\documentclass[11pt]{article}

\AtBeginDocument{%
  }

\newcommand{\bx}{\mathbf{x}}
\newcommand{\by}{\mathbf{y}}
\newcommand{\calD}{\mathcal{D}}
\newcommand{\calR}{\mathcal{R}}
\newcommand{\arb}{\mathbb{A}}
\newcommand{\PD}{\texttt{PD}}

\newtheorem{definition}{Definition}
\usepackage{multirow} 

\usepackage[usenames,dvipsnames]{xcolor}
\usepackage{bbm}
\usepackage[numbers]{natbib}
\usepackage{amsmath}
\usepackage{amsfonts}
\usepackage{amssymb}
\usepackage{booktabs}
\usepackage{graphicx}
\usepackage{relsize}
\usepackage{url}
\usepackage{hyperref}
\usepackage[margin=1in]{geometry}
\usepackage{adjustbox}
\usepackage{float}
\usepackage[caption = false]{subfig}

\newcommand{\CO}[1]{{\color{Green}#1}}
\newcommand{\CT}[1]{{\color{Blue}{#1}}}
\newcommand{\quotes}[1]{``#1''}

\begin{document}

\title{Algorithmic Arbitrariness in Content Moderation }

\author{Juan Felipe Gomez \thanks{Equal contribution,  alphabetical order.}~\thanks{ J. F. Gomez is with the Department of Physics, Harvard University (email: { juangomez@g.harvard.edu})}~, Caio Vieira Machado${}^*$~\thanks{C. Machado is with the Centre for Socio-Legal Studies, University of Oxford and Universidade de São Paulo Law School. Work done as a Research Fellow at Harvard.  (email: caio.machado@wolfson.ox.ac.uk)}~, Lucas Monteiro Paes${}^*$~\thanks{L. M.  Paes  and F. P. Calmon are with the School of Engineering and Applied Science, Harvard University (emails: { lucaspaes@g.harvard.edu,  flavio@seas.harvard.edu})}~, Flavio P. Calmon ${}^\ddag$ \thanks{Principal Investigator}}
\date{}

\maketitle
\begin{abstract}
    Machine learning (ML) is widely used to moderate online content. Despite its scalability relative to human moderation, the use of ML introduces unique challenges to content moderation. One such challenge is predictive multiplicity: multiple competing models for content classification may perform equally well on average, yet assign conflicting predictions to the same content. This multiplicity can result from seemingly innocuous choices during model development, such as random seed selection for parameter initialization. We experimentally demonstrate how content moderation tools can arbitrarily classify samples as toxic, leading to arbitrary restrictions on speech. We discuss these findings in terms of human rights set out by the International Covenant on Civil and Political Rights (ICCPR), namely freedom of expression, non-discrimination, and procedural justice. We analyze (i) the extent of predictive multiplicity among state-of-the-art LLMs used for detecting toxic content; (ii) the disparate impact of this arbitrariness across social groups; and (iii) how model multiplicity compares to unambiguous human classifications. Our findings indicate that the up-scaled algorithmic moderation risks legitimizing an \quotes{algorithmic leviathan}, where an algorithm disproportionately manages human rights. To mitigate such risks, our study underscores the need to identify and increase the transparency of arbitrariness in content moderation applications. Since algorithmic content moderation is being fueled by pressing social concerns, such as disinformation and hate speech, our discussion on harms raises concerns relevant to policy debates. Our findings also contribute to content moderation and intermediary liability laws being discussed and passed in many countries, such as the Digital Services Act in the European Union, the Online Safety Act in the United Kingdom, and the Fake News Bill in Brazil.
\end{abstract}

\section{{Introduction}}

Algorithmic Content Moderation is at the crossroads of two major challenges. First, there are growing legal, economic, and social pressures on companies to employ automated solutions for managing the vast quantities of undesirable online speech. Automated moderation tools pose new risks to individuals’ ability to express themselves and to seek out information. Second, the increasing reliance on Machine Learning (ML) models to flag undesirable speech can inadvertently lead to harmful outcomes stemming from the inherent technical limitations of ML. A key concern is \emph{predictive multiplicity} \cite{Marx2020}: competing models with similar average performance can produce conflicting individual predictions. Predictive multiplicity captures  \emph{arbitrariness} in ML model development, where seemingly innocuous choices made during training, such as the random seed used for parameter initialization, can affect individual prediction outcomes. Predictive multiplicity has been recently documented in a range of classification and prediction tasks  \cite{Hsu22, watson22} and can lead to disparate treatment of individual data points \cite{Creel2021, black2022model, Marx2020}. In this paper, we delve into the impact of predictive multiplicity and ensuing arbitrariness in content moderation, focusing on its potentially harmful impacts on freedom of expression, discrimination, and procedural fairness.

\paragraph{Context, relevance, and objectives}
The delegation of resource access and rights management to algorithms is a growing concern in the law and policy literature \cite{BKC_AIHR,Mittelstadt,wachter_right_2019}. This issue is particularly problematic in scenarios described as \quotes{algorithmic leviathans}, a term introduced by K\"onig \cite{konig2020dissecting} and adopted by Creel and Hellman \cite{Creel2021}, where algorithms excessively control the exercise of freedoms and access to resources. Our research specifically examines ML in content moderation, a critical instance of this potential \quotes{algorithmic leviathan}, where models are tasked with moderating content generated by billions of users worldwide. ML-based content moderation occurs with limited accountability, with platforms'  moderation policies often applied indiscriminately across jurisdictions. For instance companies' policies and their applications are inconsistent regarding definitions, tools, and policies \cite{Gorwa20, EFFContentMod}. 

\textbf{Algorithmic content moderation} is the application of algorithmic systems to classify user-generated content, leading to governance decisions such as content removal, geoblocking, or account takedowns \citep{Gorwa20}. While content moderation has traditionally been an industrial practice, algorithm-driven approaches adopted by social media platforms have come under increased scrutiny due to economic, social, and legal factors, such as COVID-19 disinformation and online extremism. These societal dynamics have prompted substantial legislative changes globally, ushering in new regulatory frameworks for online and third-party content \cite{Machado, SantosPush}. 

Though algorithmic content moderation is not always directly regulated, regulatory changes have increased pressure on companies to expedite content moderation through AI models. Examples of legislative shifts include the European Digital Services Act (DSA) \cite{DSA}, which adopts a risk-based approach for Very Large Online Platforms \cite{Buiten}, and Germany's NetzDG law \cite{NetzDG}, which mandates rapid content removal with minimal human oversight. Another remarkable example is the 2022 resolution passed by Brazil's Superior Electoral Court (TSE) which implemented a stringent 1-hour content removal window during the second round of the Brazilian Presidential elections, significantly increasing pressure for  up-scaling  algorithmic moderation \cite{ResolucaoTSE}. 
Content removal is also an ongoing debate in the United States. Federal laws regarding platforms' duties on third-party content are under intense debate \cite{TrumpOrder,KnoxBiden}, with states such as Florida enacting their own laws on online content governance \cite{FloridaSenate}. Though algorithmic content moderation may not be a direct target of these legislative efforts, they are nevertheless impacted by such policy changes.

Content governance on social media platforms can inherit limitations intrinsic to ML  models. We focus on one critical limitation: predictive multiplicity and the ensuing arbitrariness in models that classify toxic content. We demonstrate that predictive multiplicity is rampant in state-of-the-art language models used for toxic text classification: multiple models can achieve similar average accuracy yet conflict in classifying individual text samples as toxic. This algorithmic arbitrariness results from seemingly innocuous but impactful choices made during the development process, such as the choice of a random seed for initializing a model and parameters of differentially private training \cite{semenova2023a, Kulynych23}. These technical choices lead to outcomes that lack consistency, predictability, and adherence to established principles or logic \cite{creel_hellman_2022}. Note that we will use the term arbitrariness here in this sense (i.e., how unjustified choices in model development may lead to conflicting individual-level predictions), which might not fully correspond to other notions of arbitrariness in the ethical or legal field.

\paragraph{Research Questions and Main Contributions:}
We explore the role of predictive multiplicity in algorithmic content moderation. Our main research questions (RQs) are:
\begin{enumerate}
    \item \textbf{RQ1 - Model Disagreement:} \emph{What is the extent of disagreement in state-of-the-art ML models fine-tuned to classify toxic content?} We analyze how different models classify and conflict on the same content,  highlighting inconsistencies and potential biases in model predictions. %
    
    \item \textbf{RQ2 - Impact of Arbitrariness:} \emph{What are the varying impacts of arbitrariness across toxicity detection models on content targeting different social groups?} We investigate whether the arbitrary elements in ML model development result in disparate impacts on textual content related to different demographics, potentially leading to biases against certain groups. %
    
    \item \textbf{RQ3 - Forms of Harm:} \emph{What forms of harm stem from the results of RQ1 and RQ2?} We assess the broader implications of model disagreements and arbitrariness, their impact on freedom of expression, and describe ensuing societal and individual-level harms.
\end{enumerate}

We address the above research questions by analyzing state-of-the-art models for toxic (textual) content classification. 
Our research results are based on large language models (LLMs) fine-tuned for toxicity detection, specifically, the ToxDectRoberta \cite{ToxDectRoBERTa} model fine-tuned on the Toxigen \cite{toxigen} dataset and the RoBERTa base \cite{RobertaBase} model fine-tuned on the Jigsaw \cite{Jigsaw18, Jigsaw19, Jigsaw20} dataset. 
Our main contributions are:

\begin{itemize}
    \item We find that arbitrary decisions are rampant in LLMs fine-tuned for content moderation.  In our experiments, approximately $30\%$ of samples receive moderation decisions that can change by varying the random seed used to initialize training (i.e., LLM fine-tuning). Our results illustrate how arbitrary decisions in model development influence prediction outcomes in content moderation (Table \ref{tab:disagreementtoxic_50}). 
    \item We argue that predictive multiplicity poses a selective break from a rule-based approach to content moderation --- which should be based on the Law and content moderation policies --- and infringes upon procedural fairness. Multiplicity in algorithmic content moderation can unduly restrict individual and collective rights to freedom of expression via a random or unjustified model selection procedure.
    \item We show how predictive multiplicity and, consequently, arbitrary content moderation decisions are unequally distributed across different demographic groups targeted by the text being moderated. The incidence of arbitrary decisions can be discriminatory (Figure \ref{fig:ArbiPerGroup}). In our experiments, fine-tuned LLMs assign a higher rate of arbitrary predictions to textual content that mentions LGBTQ-related topics relative to textual content tagged as misogynistic or misandrist.
    \item Finally, aiming to understand the source of predictive multiplicity, we compare the arbitrariness in fine-tuned LLMs to human annotators. We show that models can often disagree in examples where human annotators unanimously agree should (or should not) be moderated (Figure \ref{fig:HumanVsModel}). This result demonstrates that i) there are content moderation decisions obvious to humans where ML models disagree, and ii) ML models can introduce additional arbitrariness to content moderation.
\end{itemize}

\subsection{Related Work}
\label{sec:related_work}
\paragraph{Predictive Multiplicity as a Risk}
\citet{Marx2020} showed the prevalence of arbitrary decisions in simple classification problems using tabular data (e.g., income and recidivism prediction). In the same paper \cite{Marx2020}, the authors elaborate on the potential harms of predictive multiplicity and argue that it should be measured and reported as we measure and report test error.
Follow-up work has analyzed how to measure and report multiplicity \cite{Hsu22, watson22}, the source of such phenomenon \cite{Kulynych23, semenova2023a, heljakka2023disentangling,long2023individual}, and its inevitability \cite{Lucas23}.
\citet{Creel2021} discuss the harms of predictive multiplicity and arbitrary decisions, leading the authors to adopt the term \emph{algorithmic leviathan}, initially introduced by \citet{konig2020dissecting}.
Aiming to avoid predictive multiplicity, \citet{black2022selective} proposed selective ensembles to decrease the number of conflicting predictions by taking a majority vote for the prediction across all competing models.
The work that is closest to ours is \cite{Creel2021}, which defines algorithmic arbitrariness and argues about its harms.
Our paper differs from \cite{Creel2021} by (i) focusing on specific harms of arbitrary decisions in content moderation and (ii) experimentally discovering and analyzing the harms of disparate arbitrary decisions across content targeting different demographic groups. 

\paragraph{Predictive Multiplicity as an Opportunity} 
Scholars also consider predictive multiplicity and model indeterminacy as an opportunity rather than a source of harm.
For example, multiplicity was leveraged to identify models that satisfy additional constraints beyond accuracy in \cite{Fisher19, Coston21, Xin22}.
\citet{Fisher19} used predictive multiplicity to generate model explanations.
\citet{Xin22} took advantage of predictive multiplicity to identify interpretable models. 
\citet{Coston21} developed a reduction approach to choose fairer models across all equally performing models.
This prior work clearly demonstrates the potential benefits of multiplicity. In contrast,  we focus on the potential harms when ML models used for content moderation are oblivious to predictive multiplicity and analyze the ensuing impact on freedom of expression.

\paragraph{Legal and policy aspects of content moderation} The law and policy literature on algorithmic content moderation has  focused on issues related to procedural fairness, inconsistent restriction of human rights, especially free speech, and discrimination studied by scholars such as \citet{Gorwa20, GillespieNot,GillespieContent,Douek}. We summarize these risks below. 

\textbf{Inconsistency in Moderation}: Different algorithms might produce divergent classifications for the same piece of content. 
Effectively, this means that either protected speech is being taken down or harmful speech is being tolerated. This can happen with regard to individual expressions or groups and their specific dialects. \citet{KellerOverRemove} listed a number of studies and resources that indicate the systematic over-removal of content for various reasons, including copyright infringement and toxic speech content moderation. \citet{Douek} explains that the process for how platforms' enforce their rules has shifted from a rule and proportionality-based approach to an algorithmic probability-based evaluation. The policy report produced by \citet{MixedMessage} offers a useful summary of the policy challenges around algorithmic tools for content moderation.  In particular, the report points to the issue that state-of-the-art content moderation algorithms have very limited capability of parsing meaning from text to make content moderation decisions. Algorithmic approaches also depend on clear-cut definitions of types of protected or illegal speech, which are very hard to obtain in the plurality and volume of online speech. Finally, content moderation risks having disparate impacts across social groups. Our study on multiplicity in content moderation demonstrates experimentally how arbitrariness can further increase these harms, introducing ambiguity in model outcomes, which highlights the incapacity of the models managing different meanings and contexts. Our study also identifies how multiplicity had disparate impacts across social groups, which can be an aggravating factor for already existing discriminatory inconsistencies in content moderation. In Section \ref{sec:conceptualizing-harms} we offer a legal definition of freedom of expression which we use as reference to discuss harms.

\textbf{Bias Amplification}: Expanding on the inconsistencies listed above, biased and discriminatory moderation may occur if algorithms used to moderate speech are inconsistent across different groups. For example, \citet{antonialli, Goncavles} describe how certain social groups have been targeted by overmoderation due to the dialects they use. Our work demonstrates that inconsistency and arbitrariness in algorithmic content removal can \emph{depend} on the subject of the statements being moderated and correlated with socio-demographic factors. 

\textbf{Opacity in Policy Enforcement}: Predictive multiplicity makes enforcing a consistent content policy difficult. When algorithms and company policies are opaque, it is difficult to identify the different results from competing algorithms. In this scenario, understanding which decisions align with the platform's guidelines becomes challenging. The incapacity of discerning between correct decisions, errors, and arbitrary decisions can conceal over-restriction or under-restriction of content. As \citet {GillespieContent} explains, vaguely worded terms and conditions of  services may legitimize unfair outcomes. Multiplicity can be an aggravating harm in the discussion, as the ambiguity across models is not something that is easily identified nor is it usually presented in transparency metrics and reports. 

\textbf{Lack of revision, explanation, and accountability}: 
\citet{PasqualeBlack} explains that understanding why algorithms produce conflicting decisions is crucial to correct algorithmic outcomes. In the context of algorithmic content moderation, we can only review and repair harmful moderation outcomes if we have a clear understanding of how these models are classifying statements. Here, we examine arbitrariness embedded in the process of training and fine-tuning models for content moderation affecting classification outcomes. Such instances of arbitrariness and normatively incorrect moderation are  impossible for humans to identify, accuse, and appeal by probing a  deployed  model \cite{BarocasLDA} since they depend on choices made during training, thus violating the premise that restrictions on freedom of expression need to be justified \cite{DiasTackling}.

\textbf{Conflicting Jurisdictions}: Each country has different laws regarding social media platforms. These laws uniquely impact the rights and concerns outlined. See, for example, the different approaches to intermediary liability pointed out by \citet{Machado,KellerIntermediary}, or the different legislative approaches to algorithmic discrimination, for example, referenced by \citet{wachter_why_2021,BinnsLegal} when arguing for EU or UK legal frameworks.

Based on the concerns above, in Section 2 we conceptualize the harms of arbitrariness in terms of the human rights principles freedom of expression, non-discrimination, and procedural justice to materialize the ethical debate in terms of violating specific legal entitlements. Then, in Sections 3 and 4, we experimentally investigate arbitrariness in state-of-the-art models, as outlined in our research questions. Finally, in Section 5, we interpret the harms arbitrariness according to the concepts of freedom of expression, non-discrimination, and procedural justice laid out in Section 2.
In this way, we are able to trace a causal relation between the technical phenomenon of arbitrariness and its infringement of legal values. We intentionally avoid local legislation and the granular matters of each jurisdictions to observe the overarching legal effects of arbitrariness in terms of specific international human rights principles. 

We are aware that international human rights laws and their principles are primarily applicable to states and do not directly impose obligations on private entities, including internet content companies. Each state enforces these principles within their own jurisdiction, regulating how businesses will respect these rights, and how companies should govern content in their services. We understand that nonetheless companies are directly and indirectly bound to these human right principles, either by platforms laws such as the DSA or the UK Online Safety Act, or by international frameworks and recommendations such as the UN Guiding Principles for Businesses on Human  Rights \cite{GuidingPrinciplesBusiness}.

\section{Judges Flipping Coins: Conceptualizing Harms of Arbitrariness in content moderation}
\label{sec:conceptualizing-harms}

Algorithmic arbitrariness in content moderation is a complex issue that traverses the fields of ML development, algorithmic fairness, tech policy, and  human rights law. \emph{To connect the concerns identified by this literature, we define harm associated to multiplicity as an infringement of legal principles.} We identify legal principles as standards that should be observed because they attend to notions of justice and morality \cite{dworkin_taking_2013}, and use these principles to establish a common concept of harm for the purposes of this research. To this end, we use the International Covenant on Civil and Political Rights (ICCPR) --- a widely ratified international treaty to which 173 countries are parties --- as our core legal reference \cite{ICCPR}.  Our analysis on the impact of arbitrariness on freedom of expression, non-discrimination, and procedural justice.

We chose to use International Human Rights Law because it gives us overarching global rules and common concepts to discuss the issues related to fundamental rights in content moderation \cite{DouekLimitsIHRL}. Altough the field has limitations in terms of direct applicability to national jurisdictions, International Human Rights Laws allows us to make claims related to multiplicity for content moderation that are \emph{transferable} across legal systems and can further be explored through the lens of local jurisprudence and statutes. For example, arbitrariness may cause inconsistent removal of transgender community dialects in Brazil or may cause the illegal censoring of nude art as pornography in France, which are locally forms of protected speech. These countries can use our findings to discuss causes of discriminatory moderation caused  by predictive multiplicity and develop safeguards for these issues considering local realities and rules. 

Building on the related work outlined in Section \ref{sec:related_work}, we define harm due to algorithmic arbitrariness as an infringement of three human rights and principles: Freedom of Expression, Non-Discrimination, and Procedure (including Procedural Justice). In essence, this work is not primarily focused on the \textit{errors} of these content moderation algorithms but instead identifying \textit{when and how these decision-making models }produce arbitrary outcomes.  To illustrate the harms we use an analogy, comparing a model's decision to that of a judge flipping coins to decide the outcome of a case. Though imperfect, we find this comparison makes the harms due to multiplicity more palpable, since the analogy emphasizes that the source of harm is the \emph{randomness} inherit to ML models. Next, we establish our working definition and reference for each principle.

\paragraph{Freedom of Expression}
Freedom of Expression (FoE) is defined in Article 19 of the ICCPR as:
\begin{quote} \smaller
    1. Everyone shall have the right to hold opinions without interference.
    
    2. Everyone shall have the right to freedom of expression; this right shall include freedom to seek, receive and impart information and ideas of all kinds, regardless of frontiers, either orally, in writing or in print, in the form of art, or through any other media of his choice.

    3. The exercise of the rights provided for in paragraph 2 of this article carries with it special duties and responsibilities. It may therefore be subject to certain restrictions, but these shall only be such as are provided by law and are necessary.
    (a) For respect of the rights or reputations of others;
    (b) For the protection of national security or of public order (ordre public), or of public health or morals.
\end{quote}
We interpret this rule in light of UN General Comment 34 \cite{GeneralComment34}, which emphasizes that freedom of expression is a broad and fundamental human right for realizing other human rights. It encompasses all forms of expression, including political discourse, journalism, artistic works, and religious dialogue, across various mediums like broadcasting, the internet, and public protest. The comment underscores the right to access information and recognizes the critical role of the internet and digital media in enabling and enhancing the exercise of freedom of expression, advocating for universal access to these platforms. 

This right is expansive but not absolute and, therefore, can be subject to certain restrictions. However, these restrictions must be clearly defined by law, serve a legitimate aim (such as protecting national security, public order, or the rights of others), and be necessary, proportionate, and pursue a legitimate aim. General Comment 34 \cite{GeneralComment34} explicitly denounces certain prohibitions, such as blasphemy laws and unreasonable restrictions on media, as incompatible with the ICCPR.

In the context of content moderation, the existence of predictive multiplicity in ML algorithms calls into question their ability to attend all requisites for a lawful restriction of freedom of expression. As an example, a ML model trained with random seed 1 could misapply a restriction to protected speech (e.g. journalistic speech), whereas the same model trained with random seed 42 would have correctly tolerated the statement. Such an event would be equivalent to a judge flipping a coin to decide whether the speech should be protected or taken down. This is not a contrived example: in Section \ref{sec:dataAnalysis} we observe that varying the random seed causes fine-tuned large language models to assign conflicting toxic speech predictions to 34\% of statements from a large scale dataset. 

\paragraph{Non-Discrimination}
We adopt Article 2(1) and Article 26 of the ICCPR as our definition of discrimination. They state:

\begin{quote} \smaller
\textbf{Article 2}
(1) Each State Party to the present Covenant undertakes to respect and to ensure to all individuals within its territory and subject to its jurisdiction the rights recognized in the present Covenant, without distinction of any kind, such as race, colour, sex, language, religion, political or other opinion, national or social origin, property, birth or other status.
\end{quote}

\begin{quote} \smaller
\textbf{Article 26}
    All persons are equal before the law and are entitled without any discrimination to the equal protection of the law. In this respect, the law shall prohibit any discrimination and guarantee to all persons equal and effective protection against discrimination on any ground such as race, colour, sex, language, religion, political or other opinion, national or social origin, property, birth or other status.
\end{quote}

These articles are intended to protect individuals from discrimination. We argue that ML algorithms can discriminate against specific individuals or groups for two reasons. First, these toxic statements target specific societal groups, therefore a biased under-moderation means these groups have an inferior protection from toxic speech. Second, language that is discriminatory in a broader context can be part of the dialect of a community and not offensive in that space. If this language is censored within this space, that social group is suffering over-moderation and is less able to exercise free speech. For example, this was the case identified by \citet{antonialli} with content moderation in LGBTQ discussion spaces. Based on these two observations, our experiments are able to infer the presence of discrimination by analyzing the targeted group of the statements.

Based on the discussion above, we claim that any algorithm that causes a particular individual or group to receive more or less restrictions on their speech compared to others is a discriminatory algorithm. In particular, if the magnitude of predictive multiplicity in ML algorithms is \emph{different} across groups, then such an ML algorithm is discriminatory. 

In Section \ref{sec:dataAnalysis} we experimentally observe exactly this phenomena: varying the random seed causes fine-tuned large language models to assign conflicting toxic speech predictions to 38\% of racial-based statements from a large scale dataset compared compared to 20\% of misogynistic/misandrist statements.  Such an algorithm is blatantly discriminatory, and the discrimination stems from the unequal protection of groups that are entitled to the same rights.

\paragraph{Procedural Justice} The UN Guiding Principles on Business and Human Rights \cite{GuidingPrinciplesBusiness} emphasize that businesses should identify, prevent, and mitigate human rights abuses. The human right to due process is established by Article 14(1) of the ICCPR, which states:

\begin{quote} \smaller
\textbf{Article 14}

(1) All persons shall be equal before the courts and tribunals. In the determination of any criminal charge against him, or of his rights and obligations in a suit at law, everyone shall be entitled to a fair and public hearing by a competent, independent and impartial tribunal established by law. The press and the public may be excluded from all or part of a trial for reasons of morals, public order (ordre public) or national security in a democratic society, or when the interest of the private lives of the parties so requires, or to the extent strictly necessary in the opinion of the court in special circumstances where publicity would prejudice the interests of justice; but any judgement rendered in a criminal case or in a suit at law shall be made public except where the interest of juvenile persons otherwise requires or the proceedings concern matrimonial disputes or the guardianship of children.
\end{quote}
We interpret Articles 14 and 19 (mentioned above) as jointly demanding that a restriction of a fundamental right be \textit{impartial}, \textit{fair}, and \textit{prescribed by law}. This means providing remedies through operational grievance mechanisms when harm occurs, ensuring processes are transparent and accountable.
When we translate this to ML models for content moderation, moderation needs to be \emph{explainable},  \textit{accountable}\footnote{We define an \emph{accountable} model as a model that can be understood, challenged, scrutinized, and revised. Explainability is a necessary but not sufficient condition for a model to be accountable.}, and have a \textit{rule-based} approach for limiting free speech. In this regard, the outcomes of ML models must attend converge to these legal requirements. This interpretation includes, for example, respecting the requirements from General Comment 34 \cite{GeneralComment34}(i.e. legality, necessity, proportionality, and pursuit of a legitimate aim) for restricting speech. This joint interpretation establishes the obligation of common procedural guidelines for removing speech.

The existence of predictive multiplicity in ML algorithms calls into question their ability to satisfy values of procedural justice. The \quotes{decision-making process} used by ML algorithms is fundamentally probabilistic and often random. As evidenced by the experimental observation in Section \ref{sec:dataAnalysis}, varying the random seed can dramatically alter the predictions made by fine-tuned large language models. This observation is equivalent to a judge sometimes flipping coins to determine whether to restrict or order the removal of speech. Continuing the judge analogy, the act of flipping coins to determine when to restrict speech violates procedural justice for a few reasons. First, it does not respect a rule-based approach to restricting speech,  as it is fundamentally random. Second, it is not impartial, as it is disparate across groups. Third, it is not accountable because this decision-making process is concealed. By \quotes{concealed}, we mean we cannot know if a given prediction is an instance of predictive multiplicity. In fact, this information is impossible to obtain even if we analyze the model alone, as multiplicity can only be identified when we compare predictions from multiple models. It follows that both the judge who flips coins and the ML algorithm violate procedural justice and fairness.  Since the source of the violation is randomness, this violation is \emph{independent} of the final outcome being legally correct.

\paragraph{Experimentally Measuring Harm}
To study multiplicity using the framework of legal harms we outlined above, we run multiple experiments to fine-tune various state-of-the-art models for toxic speech detection, test them across different datasets of toxic and non-toxic statements, and observe the incidence of predictive multiplicity across models, targeted groups of toxic speech, and even compare disagreement in models to disagreement in human annotation. These experiments are designed to allow us to quantitatively measure multiplicity and its potential harms, e.g. violations of FoE and procedural justice.  

Below we summarize our experimental development, which is later detailed in Section \ref{sec:mathMultiplicityBackground} and complemented in the Appendix.
We first acquire competing models by fine-tuning multiple times the RoBERTa base \cite{RobertaBase} and the ToxDectRoBERTa \cite{ToxDectRoBERTa} models for toxicity detection on large-scale datasets (with computational costs of hundreds of GPU hours) --- we use these model architectures and datasets with the goal of simulating how a company would approach content moderation.
To quantify the \textbf{extent of predictive multiplicity}, hence answering \textbf{RQ1}, we compute \emph{arbitrariness} (Definition \ref{def:Arbitrariness}) and \emph{pairwise disagreement} (Definition \ref{def:PairwseDiss}) on our competing fine-tuned models and show the prevalence of arbitrary decisions in SOTA toxicity detectors (Table \ref{tab:disagreementtoxic_50}).
Aiming to assess \textbf{how arbitrary decisions are spread across demographic groups}, answering \textbf{RQ2}, we compute arbitrariness and pairwise disagreement in sentences targeting specific social groups (Figure \ref{fig:ArbiPerGroup}). 

Next, we provide the necessary theoretical background on predictive multiplicity (Section \ref{sec:mathMultiplicityBackground}) and define the setup for the described experiments (Section \ref{sec:experimental_setup}).
Finally, in Section \ref{sec:dataAnalysis}, we display and analyze our experimental results.

\section{Background on Predictive Multiplicity}
\label{sec:mathMultiplicityBackground}
In this section, we discuss setup and notation, mathematically define the set of all competing models (Rashomon set), and define the multiplicity metrics of interest in this paper --- pairwise disagreement and arbitrariness.

\paragraph{Preliminaries}
We focus on the task of binary classification of toxic speech. Consider a dataset with $n \in \mathbb{N}$ examples $\calD \triangleq \{\bx_i, \by_i\}_{i = 1}^{n}$ where $\bx_i$ is a sentence (e.g., \quotes{I love you} and \quotes{I hate you}) and $\by_i \in \{0, 1\}$ is a binary label that is $1$ when the sentence is \quotes{Toxic} and $0$ when it is \quotes{Not Toxic}. In the open-source datasets used in this work, labels were generated by human annotators (see appendix \ref{app:datasets} for details). 
As usual, the dataset $\calD$ is partitioned into three datasets, one for training $\calD_{\texttt{train}}$, one for validation $\calD_{\texttt{val}}$, and one for testing $\calD_{\texttt{test}}$, i.e., $\calD = \calD_{\texttt{train}} \dot{\cup} \calD_{\texttt{val}} \dot{\cup} \calD_{\texttt{test}}$.
We use the training dataset ($\calD_{\texttt{train}}$) to further train (fine-tune) a machine learning model $h \in \mathcal{H}$ that takes a sentence $\bx$ and returns a binary label $h(\bx_i) \in \{0, 1\}$, $\mathcal{H}$ is the model class/architecture (e.g., all RoBERTa base \cite{RobertaBase} models with different parameters).
\footnote{Importantly, the training algorithm is \emph{random}, meaning that a different random seed initialization will yield a different trained model.} We use the validation dataset ($\calD_{\texttt{val}}$) to perform hyper-parameter tuning and performance evaluation during training, and the test dataset ($\calD_{\texttt{test}}$) for final evaluations. 

We use error to measure the quality of a model. Formally, the error of a model $h \in \mathcal{H}$ over a dataset $\mathcal{S} \subseteq \mathcal{D}$ is given by
\begin{equation}
    \text{Err}_{\mathcal{S}}(h) = \frac{1}{ \left|\mathcal{S} \right| } \sum_{\bx, \by \in \mathcal{S}} \mathbbm{1}\left [ h(\bx) \neq \by \right ],
\end{equation}
where $\mathbbm{1} \left[ \text{\emph{condition} }\right ]$ is the indicator function that outputs $1$ if   \emph{condition} is true and 0 otherwise. The training error, i.e., the error over $\mathcal{D}_\texttt{train}$, is defined as $\text{Err}_\text{train}(h)$, and similarly for testing error.

\paragraph{Competing Models and the Rashomon Effect} 
We call a fixed (e.g., deployed) model for flagging toxic content a \emph{reference model} and denote it by $h_{\texttt{ref}}$. The reference model can be, for example, the empirical risk minimizer over a training set or an already deployed model. We call the set of all models with less than $1 + \epsilon$  times the training error from $h_{\texttt{ref}}$ the \emph{Rashomon set }\cite{Fisher19, Breiman-two-cultures} and denote it by $\calR(\epsilon, h_{\texttt{ref}})$.\footnote{Reference \cite{Fisher19} defines the  Rashomon set with any arbitrary loss function evaluated on the training data. In this work, we use the 0-1 loss.}
Formally, the Rashomon set is given by: 
\begin{equation}
    \calR(\epsilon, h_{\texttt{ref}}) \triangleq \left\{ h \in \mathcal{H} \ \ | \ \ \text{Err}_\text{train}(h) \leq (1 + \epsilon) \: \text{Err}_\text{train}(h_{\texttt{ref}})  \right\},
\end{equation}
where $\epsilon$ is the \emph{Rashomon parameter} and measures how close the performance of the models is to the performance of the reference model, see \cite{Fisher19, black2022model, Hsu22, Marx2020} for related definitions. 
For the LLMs considered in this work, the Rashomon set is theoretically and computationally challenging to characterize. We resort to empirically estimating the Rashomon set via re-running the same fine-tuning pipeline with different random seeds. Each fine-tuned model gives us a sample from the Rashomon set if the model is close in performance to the reference model. We denote these \emph{Rashomon set model samples} by $\widehat{\calR}(\epsilon)$ when $h_{\texttt{ref}}$ is clear from the context.
In practice, to explore the Rashomon set, we fix a dataset $\calD_{\text{train}}$ and model architecture $\mathcal{H}$, and fine-tune as many models on $\calD_{\text{train}}$ as our computational resources allow, each time varying the random seed. We discard any models that are not within $\epsilon$ of $h_{\texttt{ref}}$ and chose $h_{\texttt{ref}}$ to be a language model freely available on HuggingFace. 

There is no standard Rashomon parameter selection method ($\epsilon$).
Most papers on predictive multiplicity resort to showing how results vary when the Rashomon parameter is changed \cite{Marx2020, Hsu22, black2022selective, semenova2023a, Kulynych23}.
Recently, \citet{Lucas23} proposed a principled manner of choosing the Rashomon parameter based on Clopper-Pearson confidence intervals. This approach --- which we refer to as the CP method --- selects $\epsilon$ based on a confidence parameter, dataset size, and the error of the reference model. We follow their approach using a confidence parameter of $95\%$ for a conservative analysis.
We also explore different confidence values in appendix \ref{sec:RashomonParameterVariation}.

\paragraph{Measuring Predictive Multiplicity}
A classification problem exhibits predictive multiplicity when models in the Rashomon set assign conflicting predictions to the same data point, formally defined in \citet[Definition 2]{Marx2020}. 
To measure predictive multiplicity, we use the following two metrics: arbitrariness, which is a generalization of ambiguity \citet{Marx2020}, and pairwise disagreement \cite{black2022model, d2022underspecification}. 

While ambiguity computes the fraction of points that at least one model in the Rashomon set disagrees with the reference model ($h_{\texttt{ref}}$), arbitrariness measures the percentage of points in the dataset that receive conflicting predictions from any two models in the Rashomon set (competing models) and it is formally defined next. 

\begin{definition}[{Arbitrariness}] The arbitrariness on a set of inputs $\mathcal{S} = \{\bx_1, ..., \bx_n\}\subseteq \mathcal{D}  $ over the Rashomon set model samples $\widehat{\calR}(\epsilon, h_{\texttt{ref}})$ is the proportion of inputs in the set $\mathcal{S}$ that receive conflicting predictions from any two models in the Rashomon set model samples:
\begin{equation}
    \widehat{\arb}(\epsilon) \triangleq \frac{1}{n} \sum_{i = 1}^{n} \mathbbm{1}[ \exists h_1, h_2 \in \widehat{\calR}(\epsilon) | h_1(\bx_i) \neq h_2(\bx_i)].
\end{equation}
\label{def:Arbitrariness}
\end{definition}

Pairwise disagreement is a per-sample measure that approximates the fraction of models in the Rashomon set that disagree on a particular prediction.
Formally, pairwise disagreement is defined as follows.

\begin{definition}[{Pairwise Disagreement \cite{black2022model, d2022underspecification}}] 
\label{def:PairwseDiss}
The pairwise disagreement for a given input $\bx \in \mathcal{D}$ over the Rashomon set model samples $\widehat{\calR}(\epsilon, h_{\texttt{ref}})$ is the proportion of pairs of models that disagree on the given input:
\begin{equation}
    \widehat{\PD}_{\epsilon}(\bx) \triangleq \frac{1}{M (M-1)} \sum_{h_i, h_j \in \widehat{\calR}(\epsilon)} \mathbbm{1}[h_i(\bx) \neq h_j(\bx)],
\end{equation}
where $M = \lvert \widehat{\calR}(\epsilon) \rvert$, i.e., $M$ is the number of models we sample from the Rashomon set via retraining.
Throughout this paper, we will report the average pairwise disagreement, given by averaging the empirical pairwise disagreement across all points in a dataset.
Formally, given a set of inputs $\mathcal{S} = \{\bx_1, ..., \bx_n \} \subseteq \mathcal{D} $ the average pairwise disagreement is given by:
\begin{equation}
    \overline{\PD}(\epsilon) \triangleq \frac{1}{n} \sum_{i = 1}^{n} \widehat{\PD}_{\epsilon}(\bx_i).
\end{equation}
\end{definition}

Arbitrariness and pairwise disagreement are both defined as a point-wise estimate over the Rashomon set samples.
To account for the error due to sampling in a finite dataset instead of using the true data distribution, we also report $95\%$ confidence intervals to their estimates using the bootstrap method from Seaborn \cite{seaborn} across the available dataset --- see Figures \ref{fig:ArbiPerGroup} and \ref{fig:HumanVsModel} for an example.

We select the above metrics because they quantify two important aspects of predictive multiplicity: (i) the fraction of samples in a dataset for which predictions are arbitrary (Defn. \ref{def:Arbitrariness}), in that a competing model would have assigned a different prediction,  and (ii) the extent to which models disagree on individual (Defn. \ref{def:PairwseDiss}). 

Given a set of models sampled from the Rashomon Set (e.g., by varying random seeds), we quantify predictive multiplicity in two steps.
First, we measure the number of arbitrary decisions (arbitrariness) made by competing models. Here, arbitrariness captures \emph{how many moderation decisions were not rule-based but just a consequence of random seed selection.} As discussed in Section \ref{sec:conceptualizing-harms}, such random decisions go against procedural fairness because they violate due process, are not \emph{accountable}, and, if the magnitude of arbitrariness is different across groups, then the impact of randomness is also \emph{disparate}. Second, we compute pairwise disagreement to estimate the number of models that disagree on their predictions.
If the number of conflicting predictions was, on average, negligible, one might argue that ignoring this conflicting minority is acceptable \cite{black2022selective}.
However, our experimental results show that such disagreement is high (Table \ref{tab:disagreementtoxic_50}), especially in specific targeted demographic groups (Figure \ref{fig:ArbiPerGroup}).
In the next section, we apply this measurement pipeline to state-of-the-art toxic text detectors.

\section{Experimental Setup}
\label{sec:experimental_setup}
This section outlines the datasets, ML models, and methodology used for evaluating predictive multiplicity in content moderation. Our goal is to describe our overall experimental approach and provide a rationale for the choice of datasets and base LLM models.

Our experiments involve \emph{fine-tuning} state-of-the-art language models on large-scale datasets. Fine-tuning refers to the act of taking a \emph{general-purpose} LLM trained on a large corpus of text, e.g. RoBERTa \cite{RobertaBase}, and further training it on a \emph{specific} objective, such as toxicity classification. Typically, this training is shorter (fewer epochs) and less intense (smaller learning rate, less updated layers) than the original training (commonly called pre-training) --- which is what motivates the term \emph{fine-tuning}. All language models referred to in this section have been fine-tuned for toxicity classification, meaning they take as input a piece of text and output either 0, denoting a non-toxic rating, or a 1, denoting a toxic rating.

\paragraph{On state-of-the-art model selection} Our first goal is to identify the state-of-the-art open-source language models that have been fine-tuned for toxicity detection. We begin by evaluating the performance of all Hugging Face \cite{huggingface} toxicity-detection language models with more than $3000$ downloads. As of January 1st, 2024, this results in 8 models (see Appendix \ref{app:model_choice}). The best-performing model (see Table \ref{tab:AccuracyHF}) was \texttt{tomh TR}\cite{toxigen},  which we will refer to as ToxiGen-RoBERTa. This model is the ToxDectRoBERTa \cite{ToxDectRoBERTa} model fine-tuned on the ToxiGen dataset \cite{toxigen}. We fix ToxiGen-RoBERTa as our reference model. Our second goal is to create competing models to ToxiGen-RoBERTa, which we did by taking the base model architecture (ToxDectRoBERTa) and \emph{fine-tuning} the model 40 times on the ToxiGen dataset while only varying the random seed between each run.\footnote{The random seed determines the weight initialization of the classification head of the language model and the shuffling of the training data, both of which lead to a different model after fine-tuning.} See Appendix \ref{app:hp} for details on the fine-tuning procedure.  We then discard the models that are worse than the reference model using the CP method from \cite{Lucas23} outlined in Section \ref{sec:mathMultiplicityBackground}, using a confidence of 95\%. This choice enables a
conservative estimate of the size of the Rashomon set and, therefore, of multiplicity across datasets. This results in a Rashomon parameter of $\epsilon = 0.016$, and us keeping 35 of the 40 models as Rashomon set samples ($\widehat{\calR}(\epsilon)$).

\paragraph{On dataset selection} Next, we used these 35 models to quantitatively measure predictive multiplicity across datasets and social groups. We use the publicly available datasets: ToxiGen \cite{toxigen}, DynaHate \cite{Dynahate}, SBF (Social Bias Frames) \cite{SBF}, HateExplain \cite{Hateexplain}, MHS (Measuring Hate Speech) \cite{MHS}, and WikiDetox \cite{WikiDetox}.
These datasets were chosen for two main reasons. 
These datasets were purposefully designed to challenge ML-based toxic text classification.
For example, ToxiGen and SocialBiasFrames (SBF) contain mostly \quotes{implicit} toxic speech \cite{toxigen, SBF}. 
DynaHate uses a human-and-model-in-the-loop process to generate a dataset designed to challenge ML models.
Second, these datasets have labels for demographic groups targeted by the text. We 
use this information to quantify and compare Arbitrariness and Pairwise Disagreement across different targeted groups (Figure \ref{fig:ArbiPerGroup}).
We also use the Measuring Hate speech (MHS) \cite{MHS} and the WikiDetox \cite{WikiDetox} datasets. 
We chose these datasets because they add one additional dimension to our analysis: the labels of multiple human annotators who detected toxicity for the sentences in the dataset.
This information enables us to compare human annotators' disagreement with model disagreement (Figure \ref{fig:HumanVsModel}).
See Appendix \ref{app:datasets} for further details on these datasets.

\paragraph{Further model selection} Moreover, we repeat the multiplicity experiment outlined above with the second-best-performing model from HuggingFace to guarantee that our experimental results are not a mere artifact of model architecture or training data selection.
This model is \texttt{s-nlp RTC}\cite{dale2021skoltechnlp}, which we will refer to as RoBERTa-Toxicity-Classifier from here on. This model is a base RoBERTa model \cite{RobertaBase} fine-tuned on the Jigsaw dataset \cite{Jigsaw18,Jigsaw19, Jigsaw20}. Due to computational limitations, we fine-tune this model 20 times and use the same CP method outlined above to discard the worst-performing models with a confidence of 95\%. This results in a Rashomon parameter of $\epsilon = 0.002$, and us keeping 16 of the 20 models fine-tuned models.  
\begin{table}[t]
  \centering
  \caption{Test accuracy for all Hugging Face toxicity detection models with more than 3k downloads and ToxiGen across different datasets.
  The best-performing model accuracy is shown in \CO{green} and the second best in \CT{blue}. See Table \ref{tab:HFModels} for the full list of selected models along with their references.}
  \label{tab:AccuracyHF}
  \begin{tabular}{lcccc}
    \toprule
    Models & Toxigen & DynaHate & SBF & HateExplain \\
    \midrule
    \texttt{martin-ha TCM} \cite{Martin} & $56.2 \%  \pm 3.5\%$ & $52.9 \% \pm 1.6\%$ & $56.5 \% \pm 1.4\%$ & $55.5 \pm 2.2\%$ \\
    \texttt{unitary TB} \cite{Detoxify} & $62.5 \% \pm 3.4\%$ & $55.2 \% \pm 1.6\%$ & $58.2 \% \pm 1.4\%$ & $64.1 \pm 2.2\%$\\
    \texttt{s-nlp RTC} \cite{dale2021skoltechnlp} & \CT{$66.9 \% \pm 3.3\%$} & \CT{$56.9 \% \pm 1.6\%$} & $62.4 \% \pm 1.3\%$ & \CT{$65.9 \pm 2.1\%$} \\
    \texttt{mohsenfayyaz TC} \cite{MohToxic} & $63.2 \% \pm 3.4\%$ & $56.1 \% \pm 1.6\%$ & \CO{$68.5 \% \pm 1.3\%$} & $63.8 \pm 2.1\%$ \\
    \texttt{unitary UTR} \cite{Detoxify} & $64.5 \% \pm 3.3\%$ & $54.6 \% \pm 1.6\%$ & $58.4 \% \pm 1.4\%$ & $65.8 \pm 2.1\%$ \\
    \texttt{nicholasKluge TM} \cite{nicholas22aira} & $58.5 \% \pm 3.5\%$& $55.2 \% \pm 1.6\%$ & $56.3 \% \pm 2.2\%$ & $62.4 \pm 2.1\%$\\
    \texttt{unitary MTXR} \cite{Detoxify} & $63.1 \% \pm 3.3\% $& $54.6 \% \pm 1.6\%$ & $60.1 \% \pm 1.4\%$ & $64.3 \pm 2.1\%$\\
    \texttt{tomh TR} \cite{toxigen} & \CO{$83.4 \% \pm 2.6\%$} & \CO{$58.1 \% \pm 1.6\%$} & \CT{$64.1 \% \pm 1.3\%$} & \CO{$67.8 \pm 2.0\%$}\\
    \bottomrule
  \end{tabular}
\end{table}

Having fine-tuned our models, in the next section, we will present how these models exhibit predictive multiplicity in accordance with the mathematical formulation in Section \ref{sec:mathMultiplicityBackground}.
For each of our findings, we also draw connections between our experimental results and their impact on principles of procedural justice, freedom of expression, and non-discrimination, based on the legal framework outlined in Section \ref{sec:conceptualizing-harms}.

\section{Data Analysis}
\label{sec:dataAnalysis}
In this section, we present our experimental results and discuss their meaning in terms of the principles defined in Section \ref{sec:conceptualizing-harms}.
As we did in Section \ref{sec:conceptualizing-harms}, we will often refer to the illustration of a judge flipping a coin to discuss the harms identified.

\subsection{Procedural Justice, Freedom of Expression, and Judges Flipping Coins}
\label{sec:numerical-procedure}

\begin{table*}[b]
  \centering
  \caption{
  Average pairwise disagreement and arbitrariness in testing time for the Toxigen fine-tuned and Jigsaw fine-tuned models in different datasets. 
  The confidence in the CP methods was chosen to be $95\%$ for a more conservative analysis.
  $95\%$ confidence intervals are shown using the standard error from the mean.
  }
  \label{tab:disagreementtoxic_50}
  \begin{tabular}{ccccc}
    \toprule
            & \multicolumn{2}{c}{Toxigen Fine-Tuned} & \multicolumn{2}{c}{Jigsaw Fine-Tuned}\\
    Dataset  & Pairwise Disagreement  & Arbitrariness  &  Pairwise Disagreement  & Arbitrariness  \\
    \midrule
    {Toxigen} & $6.8\% \pm 0.9\%$  & $28.6\% \pm 3.2\%$ & $4.3\% \pm 0.8\%$ & $15.4\% \pm 2.5\%$\\
    {DynaHate} & $8.4\% \pm 0.6\%$  & $34.1\% \pm 1.6\%$  & $6.0\% \pm 0.4\%$  & $21.8\% \pm 1.4\%$\\
    {SBF} & $8.7\% \pm 0.3\%$ & $34.4\% \pm 1.1\%$  & $7.2\% \pm 0.3\%$& $24.4\% \pm 1.0\%$\\
    {HateExplain} & $8.4\% \pm 0.6\%$  & $31.9\% \pm 2.0\%$  & $8.5\% \pm 0.6\%$ & $26.6\% \pm 2.0\%$\\
    \bottomrule
    \textbf{Total}  & $8.3\% \pm 0.2\%$  & $34.2\% \pm 0.8\%$ & $6.9\% \pm 0.2\%$ & $23.9\% \pm 0.7\%$\\
    \bottomrule
  \end{tabular}
\end{table*}

\paragraph{Technical Analysis} Our first experimental result regards the extent of arbitrariness and disagreement in our fine-tuned state-of-the-art toxicity detectors.
Table \ref{tab:disagreementtoxic_50} shows the prevalence of arbitrariness for the fine-tuned Toxigen and Jigsaw models across all tested datasets. 
We also observe that for the fine-tuned Toxigen, more than $34\%$ of all decisions made by the models at the test time are arbitrary, i.e., there exists another competing model with a conflicting prediction. For the fine-tuned Jigsaw models, this number decreases to closer to $23\%$. Moreover, both the fine-tuned Toxigen and Jigsaw models achieved a high number of conflicting predictions in the SBF dataset that contains \emph{implicit} toxic content --- which may indicate that when the toxicity is implicit, arbitrary decisions are more common.

Table \ref{tab:disagreementtoxic_50} also shows a high percentage of pairwise disagreement for the fine-tuned Toxigen and Jigsaw models across all tested datasets.
Our experiments show that using the fine-tuned Toxigen models, on average, $8.3\%$ of the pair of models disagree in their prediction --- i.e., $8.3\%$ of total pairwise disagreement.
While $6.9\%$ of the pair of models disagree for the fine-tuned Jigsaw models.
This implies that, on average, for each point that models disagree, $14\%$ of the fine-tuned Toxigen models made a prediction about sentence toxicity, and $86\%$ of the models predicted the opposite.
This high pairwise disagreement is especially relevant for methods that aim to decrease arbitrary decisions by taking a majority vote across fine-tuned competing models such as \cite{black2022selective}.

\paragraph{A Violation of Procedure and Freedom of Expression} We picture that each of our models is a judge, and each statement in the training dataset is a court case on toxic online content that they must make a ruling on. The judge's decision is binary: either take down the online post or not. Recall that the models we developed and tested are part of a Rashomon set, meaning they all have very similar accuracy and are, therefore, equally good. On average, all judges make the same number of correct rulings. However, in 34\% of court cases, at least two judges disagree on the ruling (arbitrariness). These conflicting rulings are \emph{not} a result of judges having different interpretations of the law or or having different ideologies (e.g., more or less punitive). These conflicts stem from purely random events, e.g., in 34\% of court cases the judge flips a coin to decide whether to take down the online post or not. Per Section \ref{sec:conceptualizing-harms}, such decisions are entirely detached from notions of due process, legality, and impartiality, and hence constitute a violation of procedure and freedom of expression. Bringing the discussion back to ML models, the fact that we measure a 34\% arbitrariness value due solely to \emph{random events} means these ML models, if deployed in the real world, would blatantly violate procedure and FoE (as defined in Section \ref{sec:conceptualizing-harms} ). We emphasize that if the 34\% arbitrariness value could be attributed to \emph{clear} and \emph{explainable} differences in decision-making, then this value would not be a violation of procedure and FoE. The \emph{randomness} is the source of the violation, not the magnitude of the value.

\subsection{Disparate Arbitrariness: Different Content Gets Different Coin Flips}

\begin{figure}[t]
  \centering
\subfloat[Toxigen Fine-Tuned]{
\includegraphics[width=.5\linewidth]{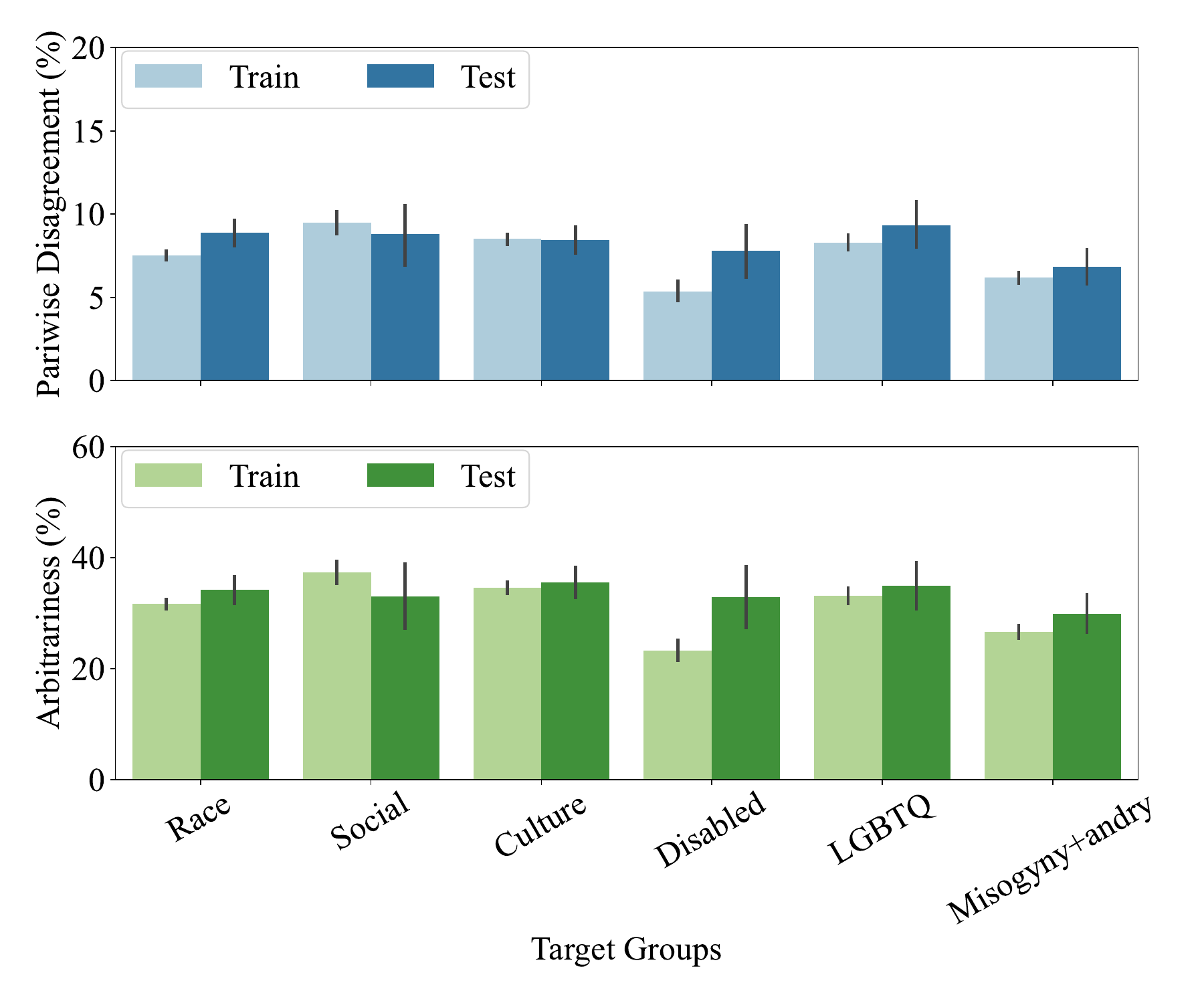}}
\subfloat[Jigsaw Fine-Tuned]{  \includegraphics[width=.5\linewidth]{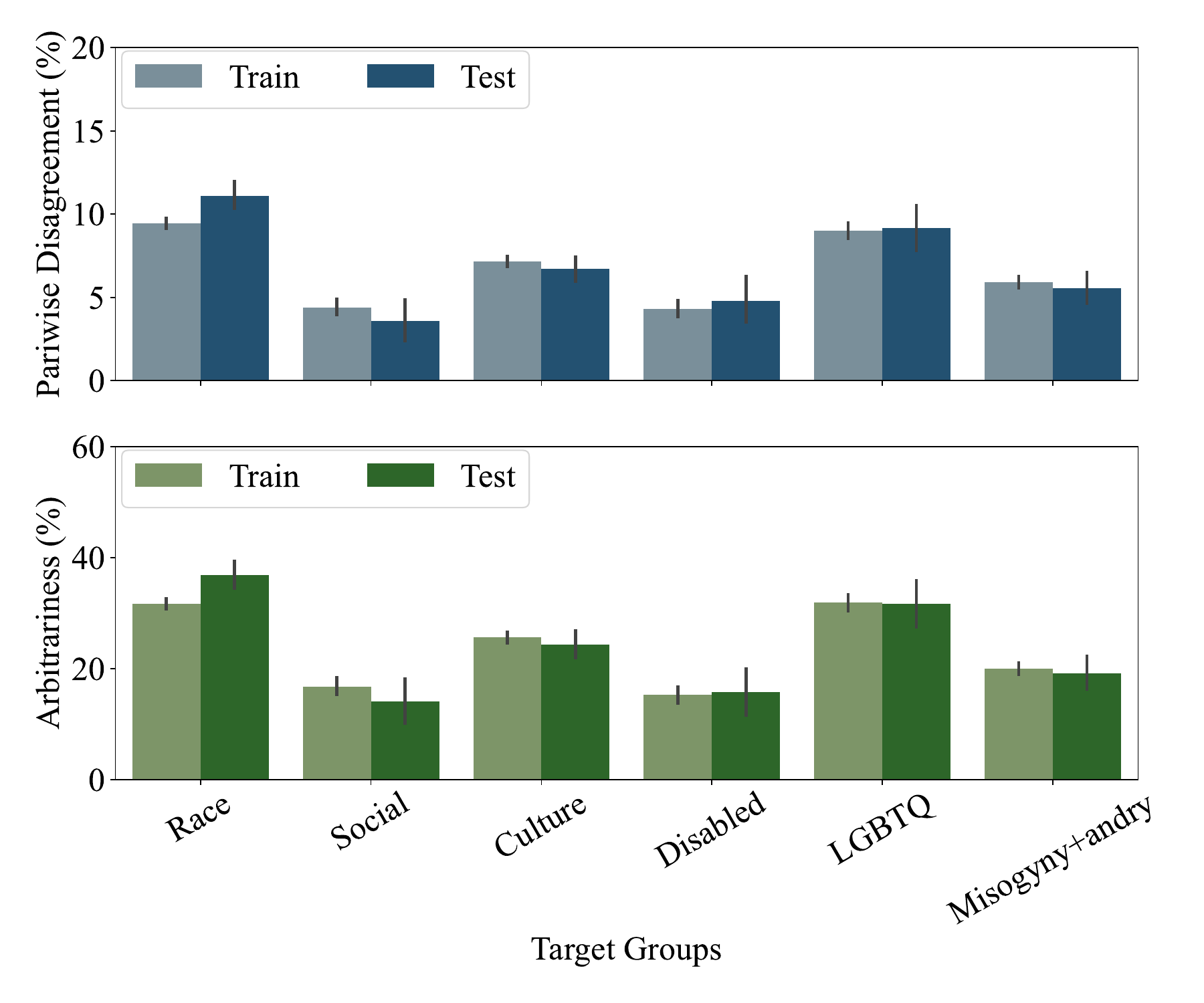}}
  \caption{Average pairwise disagreement and arbitrariness in different target groups for the fine-tuned Toxigen and Jigsaw models.
  The results show the pairwise disagreement in percentage (x-axis) for the union of four different datasets: DynaHate, SBF, Toxigen, and HateExplain.
  The results are shown for training and test partitions of each dataset.
  The confidence in the CP methods was chosen to be $95\%$.
  }
  \label{fig:ArbiPerGroup}
\end{figure}

\paragraph{Technical Analysis} Figure \ref{fig:ArbiPerGroup} indicates that the incidence of arbitrariness is not the same across all targeted groups. 
We observe that anti-LGBTQ speech consistently receives more arbitrary decisions relative to misogynist
/misandrist speech for both Toxigen and Jigsaw fine-tuned models.
Across the Toxigen fine-tuned models, anti-LGBTQ speech receives arbitrary decisions $35\%$ of the time, while misogynist/misandrist speech receives arbitrary decisions around $30\%$ of the time.
These differences are even greater on Jigsaw fine-tuned models. Moreover, racist speech has more than twice the arbitrariness of misogynist/misandrist speech on Jigsaw fine-tuned models. 
\paragraph{A Violation of Non-discrimination} Returning to the judge analogy, our experimental results indicate that decisions based on coin flips occur more frequently in certain marginalized groups than in others. 
An example would be that in 35\% of court cases concerning LGBTQ content, the judge flips a coin to decide the outcome, whereas the judge does this only 30\% of the time for misogynist and misandrist content. This unequal application of the arbitrariness based on social-demographic characteristics is a blatant violation of non-discrimination as defined in Section \ref{sec:conceptualizing-harms}. Bringing the discussion back to ML models, the fact that we measure a difference in arbitrariness values across different groups due solely to \emph{random events} means these ML models, if deployed in the real world, would violate non-discrimination. Unlike Section \ref{sec:numerical-procedure}, even if this effect could be attributed to \emph{clear} and \emph{explainable} differences in model decision-making, it would still constitute a violation of the principle of non-discrimination. People are entitled to a rule-based evaluation on whether their speech should be restricted. The uneven application of different approaches is therefore, in itself, a violation of the principle of non-discrimination. Moreover, we expand on the problems of abandoning the rule-based approach in the next sub-section.
\newline

\subsection{Comparing Human and Machine Arbitrariness: Who is Flipping Coins?}
\label{sec:human_v_model}
\begin{figure}[t]
  \centering
  \subfloat[Fine-Tuned Models Toxigen]{\includegraphics[width=.43\linewidth]{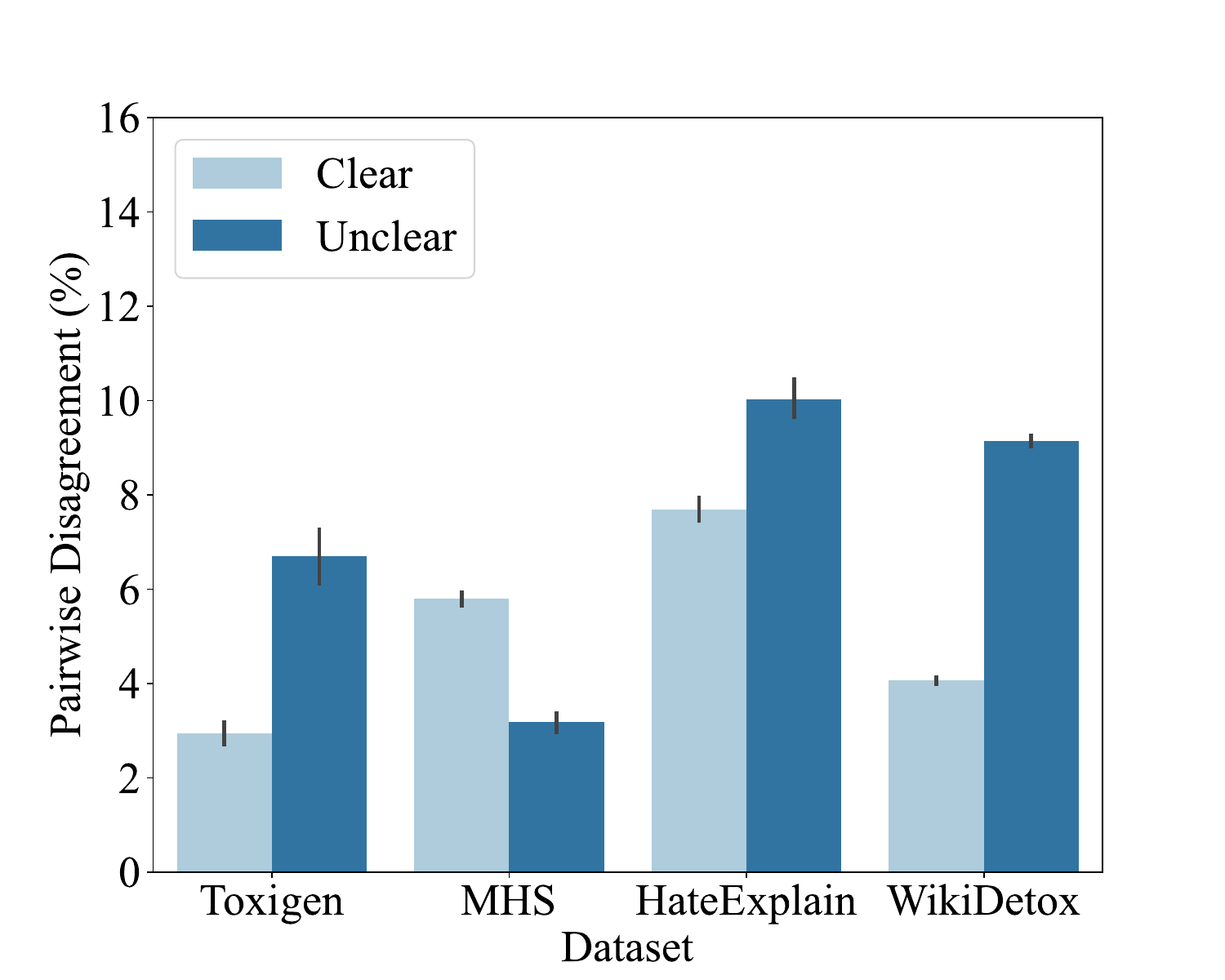}
  \includegraphics[width=.43\linewidth]{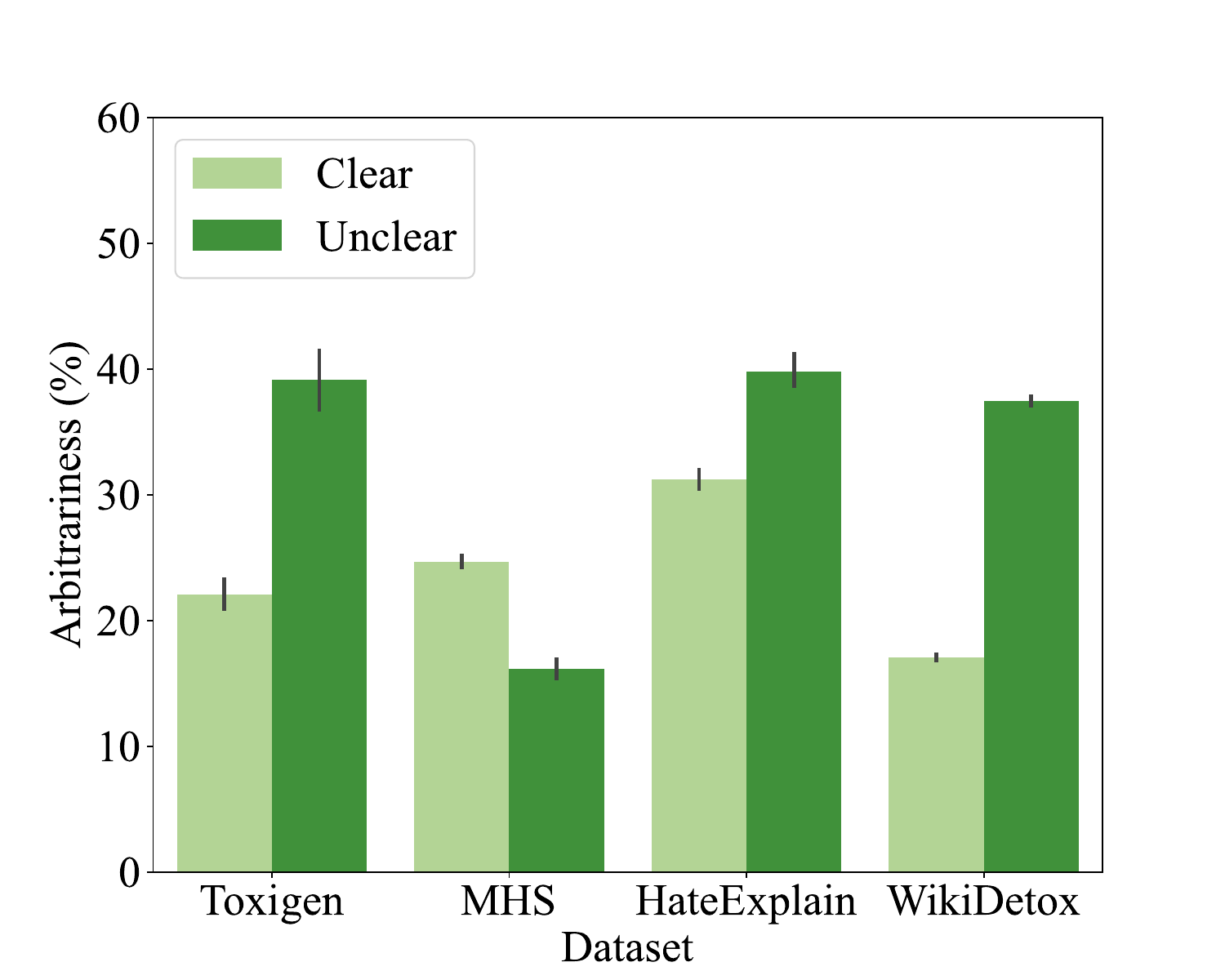}}

  \smallskip
  \vspace{-5mm}
  \subfloat[Fine-Tuned Models Jigsaw]{\includegraphics[width=.43\linewidth]{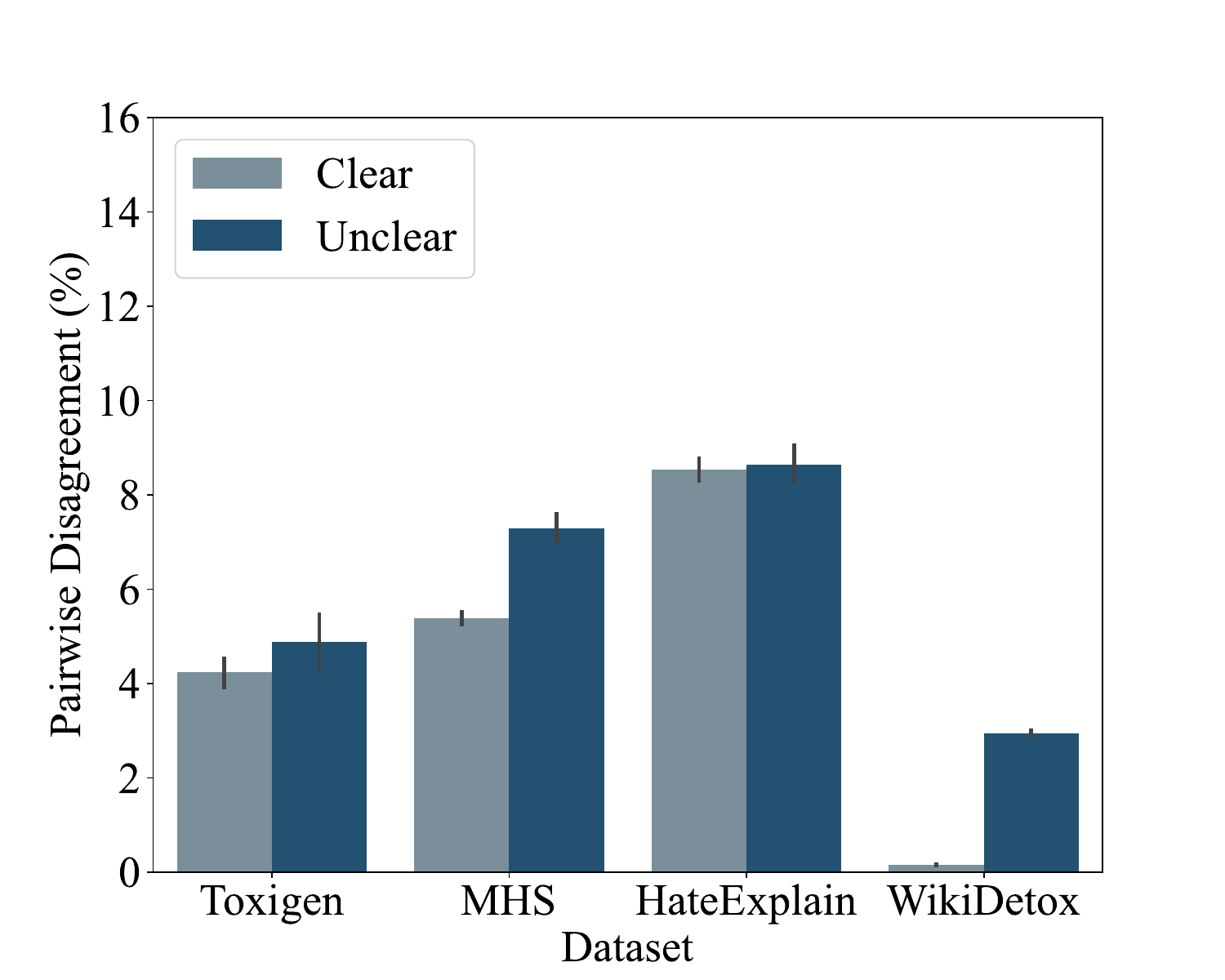}
  \includegraphics[width=.43\linewidth]{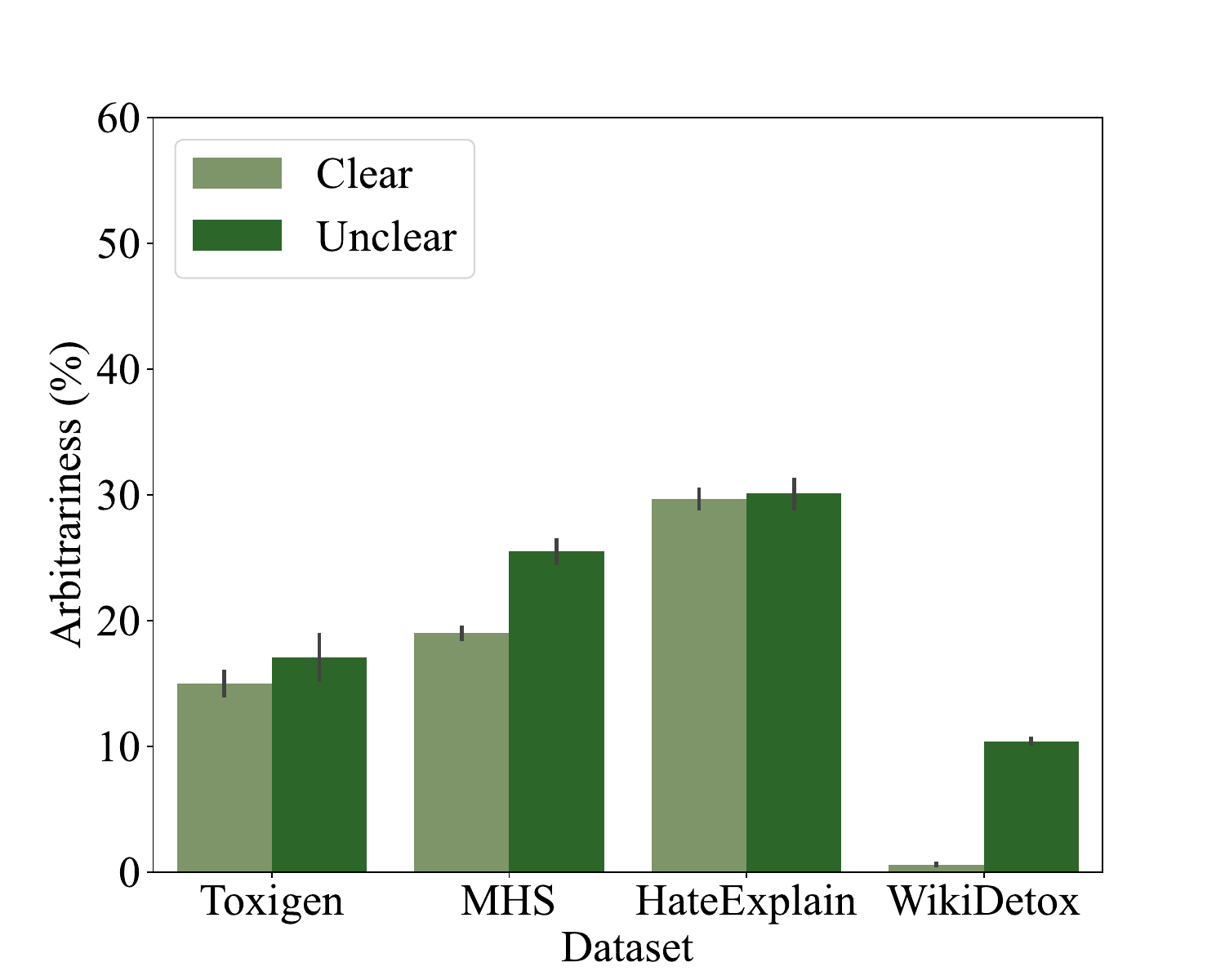}}
  \caption{Average pairwise disagreement and arbitrariness for \emph{Clear} and \emph{Unclear} sentences using the Toxigen fine-tuned and Jigsaw fine-tuned models. 
  The figure shows the pairwise disagreement estimated values along with the $95\%$ confidence intervals using the standard error from the mean.
  We consider a sentence \emph{Unclear} when at least one annotator labeled the sentence differently than others and \emph{Clear} otherwise. 
  The confidence in the CP methods was chosen to be $95\%$.
  }
  \label{fig:HumanVsModel}
\end{figure}
Finally, we compare the arbitrariness across competing ML models in the Rashomon set and across human annotators. Our goal is to verify if disagreements in predictions between fine-tuned LLMs match the disagreement observed in human annotators, in which case ML models would be replicating disagreement already present in the training data. As we see next, that is not the case.

\paragraph{Technical Analysis} From Figure \ref{fig:HumanVsModel}, we observe two  results.
\emph{First}, model disagreement tends to be higher in sentences where humans do not agree (i.e., unclear statements). 
This is an interesting finding because the models were blind to the divergence between annotators. Effectively, this implies that models, as humans, struggle with classifying certain statements.
Our \emph{second} finding is that models in the Rashomon set can display a high level of disagreement and, hence, arbitrariness in sentences in which humans \emph{unanimously} agreed on the toxicity (i.e., clear statements). In these cases, models ouput conflicting predictions when faced with evaluations that would be obvious to human annotators. Note that WikiDetox is part of the training data for the fine-tuned Jigsaw models, which is why the arbitrariness and disagreement values are noticeably small. Even in this extreme, our \emph{first} observation holds. This is further evidence that there are certain statements in these datasets that both humans and models struggle to correctly classify. 

\paragraph{A violation of rule-based approach} This point is where our analogy (un)fortunately reaches a limitation. In real life, judges will make similar decisions on most easy cases, such as an appeal for a parking ticket or the lawfulness of the social media post stating \quotes{Hello world}.
However, our findings indicate that ML models struggle over certain statements that would be obvious to any human judge, who are meant to deliver rule-based decisions. Therefore, this means that models are struggling with unambiguous statements, which raises concerns of the models' capacities to deliver good outcomes even in situations that are obvious for humans.

\section{On the Consequences of Arbitrariness}

In this section, we illustrate how selecting and deploying content moderation models at large scale, under predictive multiplicity, resembles the moral dilemma of the trolley problem. We make this comparison to  discuss the relevance of our work for law and policy decision-makers and scholars. Finally, offer insights for the path forward.

\paragraph{The inscrutable trolley problem}
The harms of arbitrariness reflect a fundamental problem for the use of ML in content moderation. ML models make statistical predictions and not interpretations of rules --- unlike human decision-makers. However, the purpose of these proxy adjudicators is to replicate the criteria of a rule-based \emph{decision-making process}, and not a probable outcome. Replacing human interpretation with statistical models to control the exercise of a right is only tolerable if they delivered similar expected outcomes, operating on the same criteria and offering procedural guarantees. We have empirically shown that is not the case. The criteria used by statistical models are often random, and oftentimes these stochastic elements are concealed from the end user.

Our work also identifies the harms stemming from arbitrary model selection (e.g., which model of the Rashomon set is chosen and deployed). When there is no clear reason for choosing one model over the other, an artificial \quotes{lottery} is created on which data points will draw the fate of being subjected to random treatment. Our results indicate that this \quotes{lottery} is not fair: different population groups targeted by the text have different likelihoods of arbitrary treatment.

If we can draw a final analogy, this creates a troubling scenario where choosing ML models is an inscrutable trolley problem. The trolley problem is a famous thought experiment in ethics and psychology involving a moral dilemma where a person must choose between actively diverting a runaway trolley to harm one person or passively allowing it to continue on its path and harm five people. Here, we do not know \emph{why} and \emph{how} companies choose between equally good models, but each one of them will cause the undue moderation of different individuals.

In the context of content moderation, from the point of view of the individuals affected by the arbitrary predictions, we effectively have a misapplication of freedom of expression rules. In other words, if arbitrariness creates harm, this is no less severe if the same prediction undid an error elsewhere. The fact that an individual suffered undue harm is a violation on its own. The fact that there was an equally possible and justifiable scenario where a different algorithm could have been deployed and preserved their rights raises questions of moral and legal responsibility. Exploring such questions in the context of specific jurisdictions is an interesting future research direction.

\paragraph{Impact on ongoing law and policy debates on content moderation}
One important debate in the platform regulation field is directly affected by these findings. It is the ongoing discussion of laws affecting content moderation, such as platform liability rules  \cite{Machado, Buiten}. Legal responsibilities imposed on service providers push companies to perform more content moderation focused on particular types of expression. Striking the right balance between free speech and expedite response, considering the volume and plurality of online communication, is a hard legal and technical task. Adding to these challenges, scientific disinformation, electoral integrity, and online extremism are all topics that have fuelled heated discussions on the need to prevent online harms while balancing international human rights  - or even questioning if international human rights are sufficient to tackle this issue \cite{DouekLimitsIHRL}.  

Our findings on predictive multiplicity increase the complexity of these tasks. Arbitrariness in algorithmic content moderation models are intrinsic characteristics of ML models with non-negligible potential for harmful outcomes, increasing the obligations companies should have in terms of accountability, transparency, and mechanisms of revision.  Moreover, as companies are assigned with increasingly complex content moderation tasks, and therefore more complex models are employed, we do not know how arbitrariness might affect outcomes. For instance, in 2022 the Brazilian Electoral Courts \cite{ResolucaoTSE} ordered the removal of content that was \quotes{similar} to content that had a previously been appreciated with a removal order. The time-frame for companies to respond, in the election periods, varied between 3 hours and 1 hour, after which the companies were subject to heavy fines. To attend these legal requirements, companies might rely on other ML models to appreciate \quotes{similarity}  at scale (whatever that might mean). 

This context of heightened reliance on models to perform highly subjective and interpretative legal tasks raises the question of what might be the outcomes of content moderation when arbitrariness across multiple models are put together. We must consider as well that these various models  are employed to analyze increasingly complex media (e.g. voice, images, and video). These concerns about arbitrariness in legally mandated content moderation can be extended to other statutes such as the UK Online Safety Act \cite{UKSafetyAct} . 

In sum, a number of laws around the world are pressuring industry to expand content-moderation tools  for various legal reasons, ranging from copyright, to public health, to national security. The consequence of these legislative and regulatory changes might be the intensified used of complex models of content moderation which are part the infrastructure governing online speech \cite{GillespieNot}. Our findings shed light on intrinsic legal complexities of these models.

\paragraph{Limitation of our work}
Some limitations of our research offer paths for future work and further development of this process. We only measured multiplicity across binary toxicity detection. However, models that predict beyond binary toxicity (e.g., models that predict the level of toxicity) may be used, affecting our results. We also didn't investigate the possibility of a statement fulfilling multiple categories of toxic speech, and we also know that content moderation may prompt different governance decisions other than simply content removal (e.g., reducing reach and labeling). Future work for a more nuanced discussion of arbitrariness in content moderation should explore these dimensions. 

\section{Conclusion}
This paper shows that predictive multiplicity is present in state-of-the-art content moderation models generating arbitrary moderation decisions, particularly in large language models for toxicity detection.
We explore the impact of this finding on prinicples of freedom of expression, non-discrimination, and procedural justice.
Then, we show that arbitrary decisions are not uniformly spread across all texts and that it is more common in texts that target specific demographic groups (e.g., anti-LGTBQ posts) and discuss the implications of this finding in terms of the principle of non-discrimination.
Finally, we check if arbitrary decisions from content moderation models align with the conflicting moderation from humans and find that the arbitrary decisions of models are also present in sentences in which human annotators unanimously moderate.

\paragraph{The path forward}
We conclude that ML models are not perfect proxies for humans when evaluating free speech. With this conclusion, we do not mean to claim that algorithmic content moderation shouldn't be used, instead we think this deployment needs to be much more nuanced and accountable. First, we need to have explainability and transparency over arbitrary decisions in the development and deployment process and analyze if the criteria used to produce the moderation decision respect the criteria we expect in terms of company policies and legal rules. Second, we need to understand how arbitrariness disparately affects subsets of the population and develop techniques to mitigate this impact.
Finally, it is also necessary to investigate a more nuanced approach to content moderation, where certain variables (e.g., thematic content, socio-demographic factors, type of illegal or harmful speech) should prompt more controls and human revision.

\section{Acknowledgements and Funding Sources}

This material is based upon work supported by the National Science Foundation under grants CAREER 1845852, CIF 1900750, CIF 2312667 and FAI 2040880. This material is based upon work supported by the U.S. Department of Energy, Office of Science, Office of Advanced Scientific Computing Research, Department of Energy Computational Science Graduate Fellowship under Award Number DE-SC0022158.
Caio Machado thanks the Coordenação de Aperfeiçoamento de Pessoal de Nível Superior - Brasil (CAPES) for partly funding this study and his visit at Harvard SEAS (Finance Code 001); and also expresses his gratitude for support provided by the Economic and Social Research Council (ESRC) Doctoral Training Parternship scheme for funding his work at the University of Oxford.

\bibliographystyle{ACM-Reference-Format}
\bibliography{references}

\appendix
\newpage

\section{Preliminaries}

In this supplementary material, we provide the following information:
\begin{itemize}
    \item Section \ref{sec:detailsOnExperimentDesign} provides further details on the datasets, Hugging Face models evaluation, fine-tuning procedures, and fine-tuned model performance.

    \item Section \ref{sec:RashomonParameterVariation} provides a further exploration of our experiment. Particularly, it shows (i) multiplicity metrics across different dataset partitions, (ii) pairwise disagreement and arbitrariness values across different datasets, and (iii) multiplicity metrics across demographics for different confidence values from the CP method.

\end{itemize}

\section{Experiment Design Details}
\label{sec:detailsOnExperimentDesign}

In this section, we provide more details on i) the datasets used, ii) the search for state-of-the-art models, iii) the used hyperparameter tuning procedure, and iv) the fine-tuned models.

\subsection{Further Dataset Information}
\label{app:datasets}

We analyze the performance of text classification models across four datasets: {ToxiGen} \cite{toxigen}, {DynaHate} \cite{Dynahate}, {SocialBiasFrames} \cite{SBF}, and {HateExplain} \cite{Hateexplain}. These datasets were chosen for several reasons. First, these are datasets purposefully designed to challenge ML-based toxic text classification. For example, ToxiGen and SocialBiasFrames contain mostly \quotes{implicit} toxic speech devoid of explicit profanity, slurs, or swearwords which could be easily flagged \cite{toxigen, SBF}. 
DynaHate uses a human-and-model-in-the-loop process to generate a dataset designed to fool ML models. Second, these datasets have labels for demographic groups targeted by the text.
This information enables us to quantify and compare Arbitrariness and Pairwise Disagreement across different target groups and report disparities in Section \ref{sec:dataAnalysis}.
In addition to these datasets, we also use the Measuring Hate speech (MHS) \cite{MHS} and the WikiDetox \cite{WikiDetox} datasets. 
We chose these datasets because they add an additional dimension to our analysis: the labels of multiple human annotators who detected toxicity in each statement in the dataset.
This information enables us to  compare human annotators' disagreement with model disagreement  in Section \ref{sec:human_v_model}.

See Table \ref{tab:Datasets_summary} for a summary of all datasets used in this work. Here, the \quotes{Unique Samples} column refers to the number of unique sentences that appear in the corresponding datasets across train, test, and validation. The \quotes{Human Annotators per Sample} column refers to the number of independent human annotators that saw each sample. For example, an entry such as \quotes{1-5} means between 1 and 5 human annotators saw every sample in the dataset.

\begin{table*}[t]
  \centering
  \caption{Summary of all datasets used.}
  \label{tab:Datasets_summary}
  \begin{tabular}{ccccc}
    \toprule
    Dataset & \% Toxic & Unique Samples & \# human annotators per sample\\
    \midrule
    ToxiGen & 42.5 & 6,514 & 3 \\
    Jigsaw & 8.1 & 2,223,061 (130,320 used) & 1-3,563\\
    DynaHate & 43.7 & 33,677 & 1-5\\
    SocialBiasFrames & 46.8 & 45,223 & 1-20 \\
    HateExplain & 59.4 & 19,229 &  3\\
    MeasuringHateSpeech & 20.5\ & 39,555 & 1-815\\
    WikiDetox & 7.7 \ & 197,578 & 8-46\\
    \bottomrule
  \end{tabular}
\end{table*}

\subsection{Model Search On HuggingFace}
\label{app:model_choice}
We include a screen shot of the HuggingFace platform listing the most downloaded language models for toxicity detection as of January 1st, 2024. The purpose of this screenshot is to keep historical proof that we tested all models with more than 3000 downloads as of the time of writing. Note that \texttt{s-nlp/russian\_toxicity\_classifier}
and \texttt{cointegrated/rubert-tiny-toxicity} are Russian language models and hence outside the scope of this paper. For the same reason, \texttt{naot97/vietnamese-toxicity-detection\_1}, a Vietnamese language model, was not considered. Moreover, \texttt{rungalileo/toxic-bert-quantized-traced} is a distilled / quantized version of \texttt{unitary/toxic-bert}, hence we opted to use only \texttt{unitary/toxic-bert}. 

See Table \ref{tab:HFModels} for the full list of selected models along with their references.

\subsection{Hyperparameters}
\label{app:hp}
The accuracy of fine-tuned language models depends heavily on a multitude of hyperparameters. In the main body, we retrain two different model types multiple times: the ToxiGen-RoBERTa \cite{toxigen} and the RoBERTa-Toxicity-Classifier \cite{dale2021skoltechnlp}. In this section, we detail the hyperparameters used in the main body. 

\begin{figure}[t]
  \centering
  
\subfloat[Toxigen Fine-Tuned]{
\includegraphics[width=.4\linewidth]{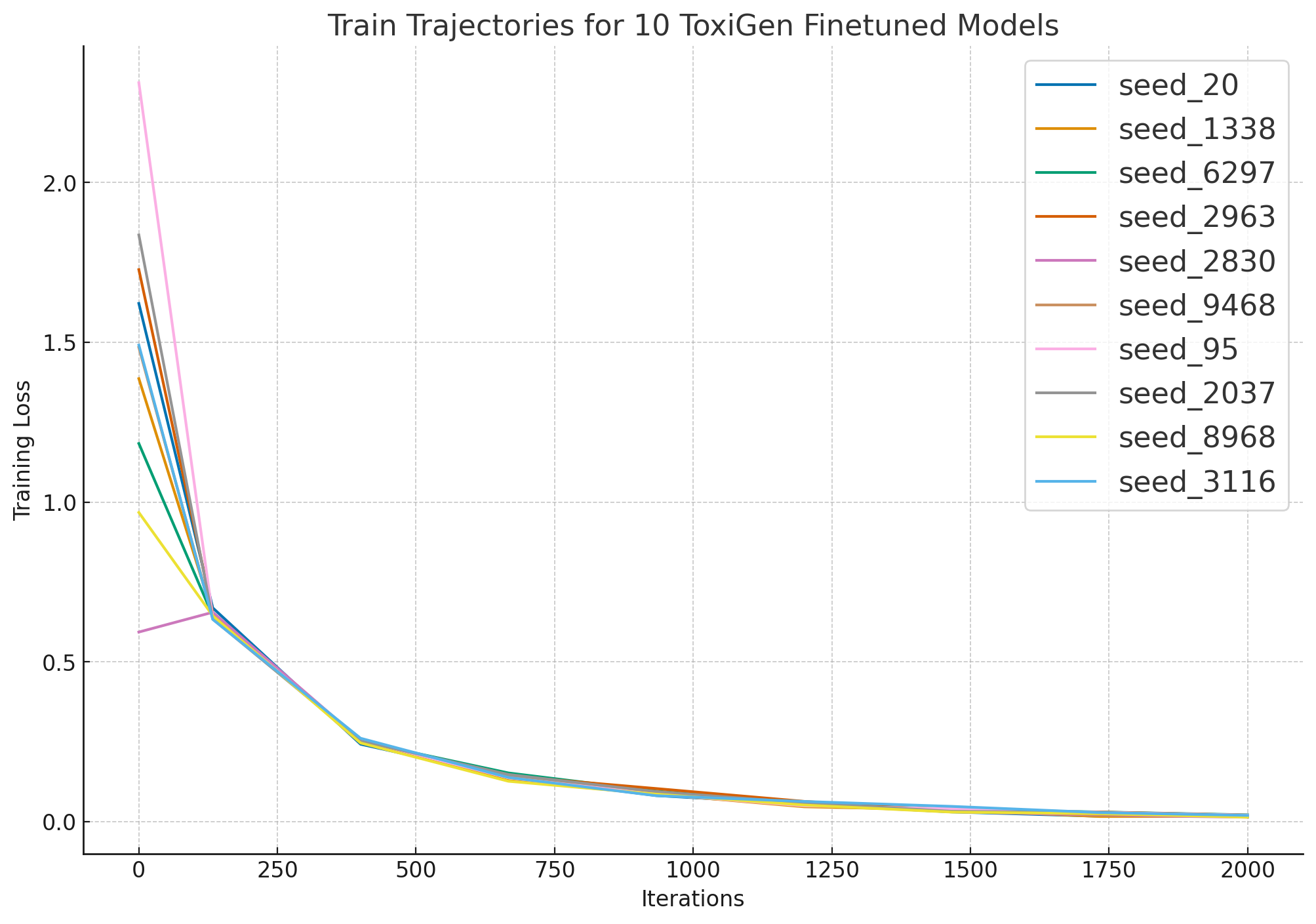}}
\subfloat[Jigsaw Fine-Tuned]{  \includegraphics[width=.4\linewidth]{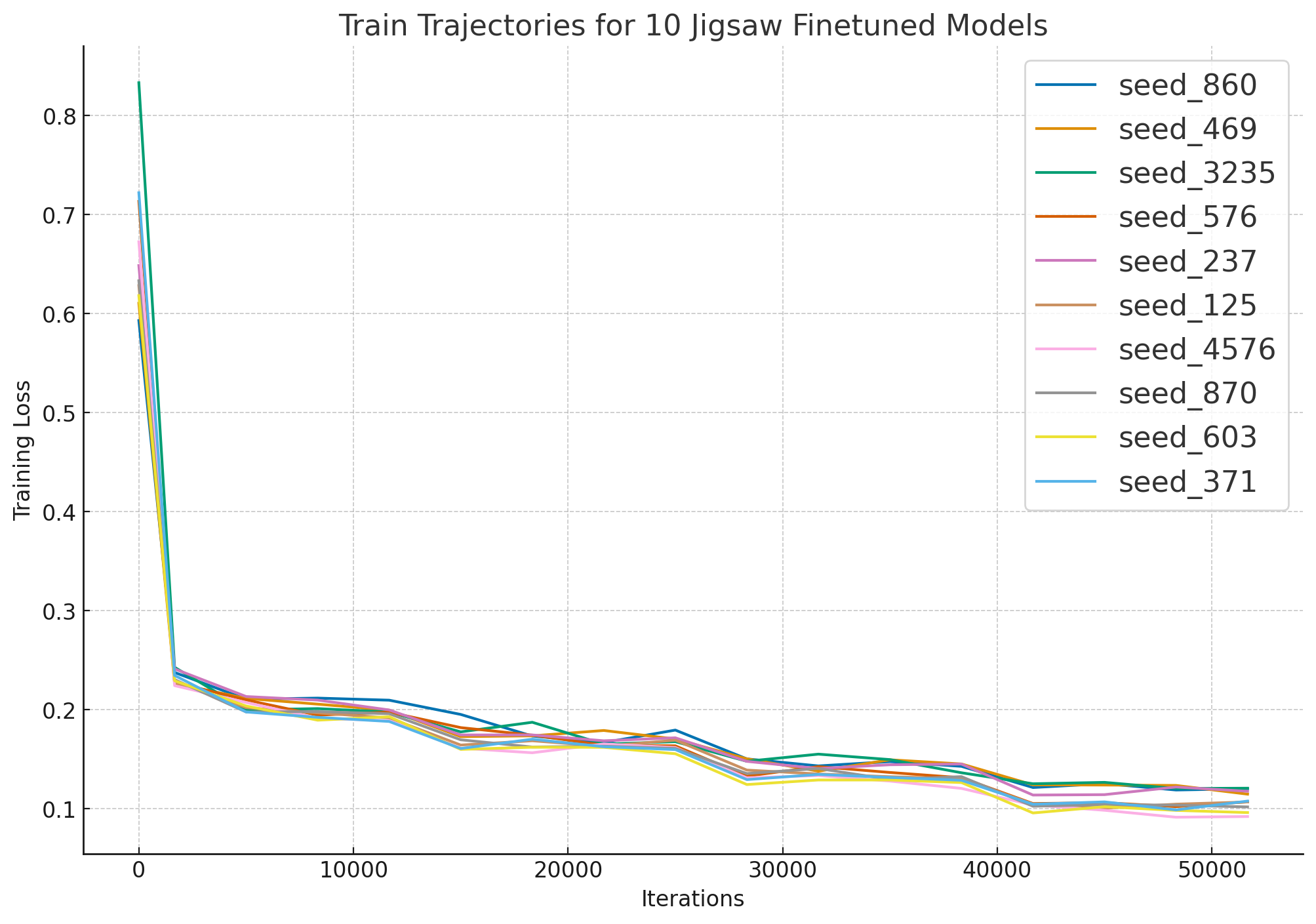}}
  \caption{Training trajectories for the fine-tuned ToxiGen and Jigsaw models over 10 randomly chosen seeds. 
  }
  \label{fig:train_traj}
\end{figure}

\paragraph{ToxiGen-RoBERTa:} Retraining the ToxiGen-RoBERTa model  was done by fine-tuning the ToxDectRoBERTa model \cite{ToxDectRoBERTa} ($\sim$ 355 million trainable parameters) on 4,601 training examples from the human annotated subset of the ToxiGen dataset \cite{toxigen}. In particular, we trained on a subset of the ToxiGen data used by \cite{Post-processed-toxigen} that removed prompts for which 3 annotators disagreed on the target group. Moreover, no quantization was done on the ToxDectRoBERTa model, and all training runs were performed on a 80Gb A100 GPU. We fixed the number of epochs to 10 and performed an extensive hyper-parameter sweep over:
\begin{itemize}
    \item learning rate: Logarithmically spaced values from $10^{-6}$ and $10^{-4}$.
    \item batchsize: Three values $\in \{8, 16, 32 \}$.
    \item Weight decay: Linearly spaced values from $0$ and $0.1$ with a 0.01 spacing.
    \item Warmup Steps: Linearly spaced values from 0 to $30\%$ of an epoch  with a $5\%$ spacing. 
\end{itemize}
All other hyperparameters were set to the default that Huggingface’s sequence classification routine uses. In particular, this means a Linear learning rate schedule with the AdamW optimzer. The sweep was done via the Trainer API from HuggingFace Transformers with the Optuna \cite{Optuna} backend, which used evaluation accuracy to prune unpromising trails early in training. In total, Optuna made 60 complete training runs (the average run took an hour and 20 minutes on an A100 GPU 80Gb). The optimal parameters were found to be: learning rate: 1e-5, batch size: 32, weight decay: 0.09, and warmup ratio: 0.1. The random seed used for the best run was 6. All ToxiGen fine tuned models (i.e., those used in the multiplicity experiments) used these hyperparameters, except for random seed. The seeds used for the ToxiGen fine tuned models were randomly generated 3 and 4 digit integers sampled using \cite{RandomOrg}. See Figure \ref{fig:train_traj} for a plot of the training trajectories of 10 of the random seeds.

\paragraph{RoBERTa-Toxicity-Classifier} Retraining the RoBERTa-Toxicity-Classifier was done by fine-tuning the base RoBERTa model \cite{RobertaBase} ($\sim$ 124 million trainable parameters) on 100,000 training examples sampled uniformly from the concatenated Jigsaw dataset \cite{Jigsaw18, Jigsaw19}. Moreover, no quantization was done on the RoBERTa model, and all training runs were performed on a 80Gb A100 GPU. In practice, the significantly larger dataset size meant that fine-tuning this RoBERTa model was approximately 3 times slower than fine-tuning the Toxigen models. Due to the increased computational cost of training these models compared to the ToxiGen models, we did not as extensive of a hyperparameter sweep. We set the batch size to 8 (for faster training time), and did a grid search for 4 epochs over four learning rates $\{10^{-6}, 10^{-5}, 2 \times 10^{-5}, 10^{-4}\}$. The best was found to be $2 \times 10^{-5}$. Then, we increased the batch size to as large as our memory allowed (32), and kept all other hyperparameters set to the default in  Huggingface’s sequence classification routine (notably: weight decay:0 and no warmup steps). All Jigsaw fine tuned models used these hyperparameters. The seeds used for the Jigsaw models were randomly generated 3 and 4 digit integers sampled using \cite{RandomOrg}. The average Jigsaw model took approximately 3 hours and 15 minutes to fine tune. See Figure \ref{fig:train_traj} for a plot of the training trajectories of 10 of the random seeds.

\begin{figure}
    \centering
    \includegraphics[width=.6\linewidth]{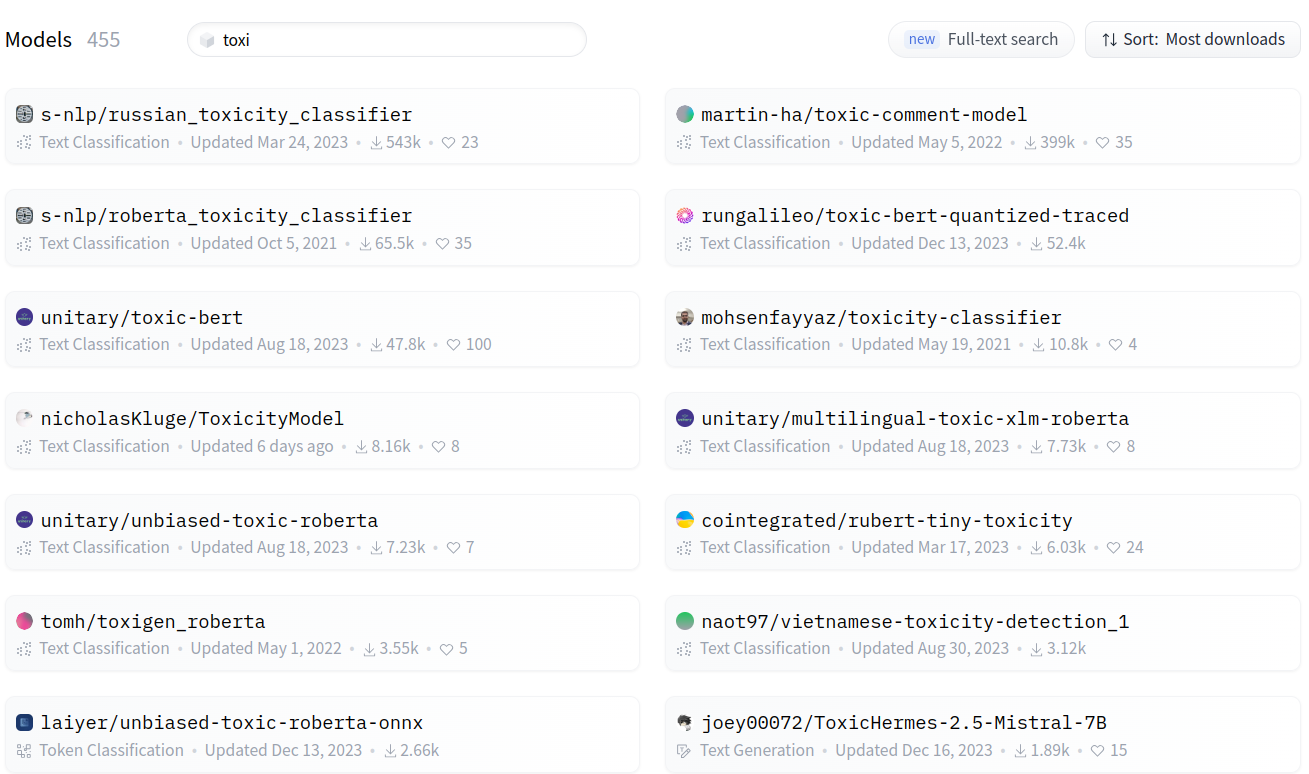}
    \caption{Screenshot of the HuggingFace platform's most popular toxic detection models as of the writing of this paper}
    \label{fig:HF-screnshot}
\end{figure}

\begin{table*}[b]
  \centering
  \caption{All considered Hugging face models.}
  \label{tab:HFModels}
  \begin{tabular}{cc}
    \toprule
    Model Name and Link &  Reference\\
    \midrule
     \href{https://huggingface.co/martin-ha/toxic-comment-model}{\texttt{martin-ha/toxic-comment-model}} & \citet{Martin}\\
    \href{https://huggingface.co/unitary/toxic-bert}{\texttt{unitary/toxic-bert}} & \citet{Detoxify}\\
    \href{https://huggingface.co/s-nlp/roberta_toxicity_classifier}{\texttt{s-nlp/roberta\_toxicity\_classifier}} &  \citet{dale2021skoltechnlp}\\
    \href{https://huggingface.co/mohsenfayyaz/toxicity-classifier}{\texttt{mohsenfayyaz/toxicity-classifier}} & \citet{MohToxic} \\
   \href{https://huggingface.co/unitary/unbiased-toxic-roberta}{ \texttt{unitary/unbiased-toxic-roberta}} & \citet{Detoxify}\\
\href{https://huggingface.co/nicholasKluge/ToxicityModel}{\texttt{nicholasKluge/ToxicityModel}} & \citet{nicholas22aira}\\
    \href{https://huggingface.co/unitary/multilingual-toxic-xlm-roberta}{\texttt{unitary/multilingual-toxic-xlm-roberta}} &  \citet{Detoxify}\\
    \href{https://huggingface.co/tomh/toxigen_roberta} {\texttt{tomh/toxigen\_roberta}} & \citet{toxigen} \\
    \bottomrule
  \end{tabular}
\end{table*}

\subsection{Fine-Tuned Models Performance}
In Table \ref{app:fineTunedPerformances}, we show the performance of the models we fine-tuned and compare it against the reference models. 
The line Reference in Table \ref{app:fineTunedPerformances} shows the accuracy of the reference ToxiGen-RoBERTa model \cite{toxigen} and RoBERTa-Toxicity-Classifier \cite{dale2021skoltechnlp} train and test accuracies.
The lines Minimum, Mean, and Maximum show the minimum, average, and maximum accuracies across all our fine-tuned models.
We observe that both the train and test performance of our models approximates the reference models deployed in Hugging-Face.
Surprisingly, the fine-tuned Jigsaw models perform as well as its reference model that was trained in 10 times more data from the same dataset.

\label{app:fineTunedPerformances}
\begin{table*}[!htbp]
  \centering
  \caption{
  Accuracy of the reference models from Hugging Face and our Fine-tuned models.
  The column Toxigen represents the accuracy of the models fine-tuned in the Toxigen dataset.
  The column Jigsaw represents the accuracy of the models fine-tuned in the Jigsaw dataset.
  The reference line shows the accuracy from the models deployed in Hugging Face.
  The lines Minimum, Mean, and Maximum show the minimum, average, and maximum accuracies across all our fine-tuned models.
  }
  \label{tab:finetunedmodelmetrics}
  \begin{tabular}{cccc}
    \toprule
    Accuracy & Data Split & Toxigen & Jigsaw \\
    \multirow{2}{*}{Reference} & Train & $96.0\%$ & $95.7\%$ \\ 
                        & Test & $83.4\%$  & $95.3\%$\\ 
\midrule
    \multirow{2}{*}{Minimum} & Train  & $94.6\%$ & $93.6\%$\\ 
                        & Test & $83.4\%$ & $92.8\%$\\ 
\midrule
    \multirow{2}{*}{Mean} & Train & $98.2\%$ & $96.6\%$\\ 
                        & Test  & $85.0\%$ & $94.1\%$\\ 
\midrule
    \multirow{2}{*}{Maximum} & Train  & $99.8\%$  & $100\%$\\ 
                        & Test & $86.8\%$ & $100\%$\\ 
    \bottomrule
  \end{tabular}
\end{table*}

\section{Further Experimental Results}
\label{sec:RashomonParameterVariation}
In this section, we show the main results in the paper for difference values for the Rashomon parameter given by the selection of confidence values for the CP method \cite{Lucas23}.
Additionally, we also show arbitrariness and pairwise disagreement across demographics for datasets. 

\subsection{Arbitrariness with Different Confidences}

We start by showing the pairwise disagreement and arbitrariness values for the testing partition of Toxigen, DynaHate, SBF, and HateExplain.
We show these results for two different confidence levels in the CP method: $50\%$ and $1\%$. When confidence is smaller, more models are considered to be in the Rahsomon set but with a higher probability of wrong model inclusion in the set.

Table \ref{tabapp:PDAr50} shows pairwise disagreement and arbitrariness for a confidence level in the CP method equal to $50\%$ and Table \ref{tabapp:PDAr1} shows results with confidence $1\%$.
We observe that, compared with Table \ref{tab:disagreementtoxic_50}, the disagreement and arbitrariness values of Tables \ref{tabapp:PDAr50} and \ref{tabapp:PDAr1} are higher as a consequence of models with higher error being included as samples of the Rashomon set.

\begin{table*}[!htbp]
  \caption{
  Average pairwise disagreement and arbitrariness in testing time for the Toxigen fine-tuned and Jigsaw fine-tuned models in different datasets. 
  The confidence in the CP methods was chosen to be $50\%$ for a more conservative analysis.
  $95\%$ confidence intervals are shown using the standard error from the mean.
  }
  \label{tabapp:PDAr50}
  \begin{tabular}{ccccc}
    \toprule
            & \multicolumn{2}{c}{Toxigen Fine-Tuned} & \multicolumn{2}{c}{Jigsaw Fine-Tuned}\\
    Dataset  & Pairwise Disagreement  & Arbitrariness  &  Pairwise Disagreement  & Arbitrariness  \\
    \midrule
    {Toxigen} & $6.8\% \pm 0.9\%$  & $28.8\% \pm 3.2\%$ & $4.5\% \pm 0.8\%$ & $16.2\% \pm 2.6\%$\\
    {DynaHate} & $8.4\% \pm 0.5\%$  & $34.3\% \pm 1.6\%$  & $6.1\% \pm 0.4\%$  & $22.7\% \pm 1.4\%$\\
    {SBF} & $8.6\% \pm 0.4\%$ & $35.4\% \pm 1.3\%$  & $7.3\% \pm 0.3\%$& $25.1\% \pm 1.0\%$\\
    {HateExplain} & $8.0\% \pm 0.6\%$  & $32.3\% \pm 2.0\%$  & $8.8\% \pm 0.2\%$ & $30.7\% \pm 2.0\%$\\
    \bottomrule
    \textbf{Total}  & $8.3\% \pm 0.2\%$  & $34.0\% \pm 0.8\%$ & $7.1\% \pm 0.2\%$ & $24.8\% \pm 0.7\%$\\
    \bottomrule
  \end{tabular}
\end{table*}

\begin{table*}[!htbp]
  \caption{
  Average pairwise disagreement and arbitrariness for the Toxigen fine-tuned and Jigsaw fine-tuned models in different datasets. 
  The confidence in the CP methods was chosen to be $1\%$, including all fine-tuned models.
  }
  \label{tabapp:PDAr1}
  \begin{tabular}{cccccc}
    \toprule
    Dataset & \multicolumn{2}{c}{Toxigen Fine-Tuned} & \multicolumn{2}{c}{Jigsaw Fine-Tuned}\\
    & \multicolumn{1}{c}{Pairwise Disagreement} & \multicolumn{1}{c}{Arbitrariness} & \multicolumn{1}{c}{Pairwise Disagreement} & \multicolumn{1}{c}{Arbitrariness} \\
    \midrule
    {Toxigen} & $6.9\% \pm 0.9\%$  & $29.6\% \pm 3.2\%$ &  $4.7\% \pm 0.8\%$ & $16.7\% \pm 2.6\%$\\
    {DynaHate} & $8.6\% \pm 0.5\%$  & $35.1\% \pm 1.6\%$ & $6.3\% \pm 0.4\%$ & $23.6\% \pm 1.4\%$\\
    {SBF} & $8.7\% \pm 0.4\%$ & $35.9\% \pm 1.3\%$ & $7.5\% \pm 0.3\%$ & $25.6\% \pm 1.0\%$\\
    {HateExplain} & $8.1\% \pm 0.6\%$ & $32.8\% \pm 2.0\%$ & $9.0\% \pm 0.6\%$ & $31.6\% \pm 2.0\%$\\
    {WikiDetox} & $6.3\% \pm 0.1\%$ & $26.5\% \pm 0.4\%$ & $1.3\% \pm 0.1\%$ & $4.7\% \pm 0.2\%$\\
    \bottomrule
    \textbf{Total} & $7.2\% \pm 0.2\%$ & $25.4\% \pm 0.7\%$ & $8.4\% \pm 0.2\%$ & $34.6\% \pm 0.3\%$\\
    \bottomrule
  \end{tabular}
\end{table*}

\subsection{Multiplicity Across Demographics}
\begin{figure}[!htbp]
  \centering
\subfloat[Toxigen Fine-Tuned]{
\includegraphics[width=.4\linewidth]{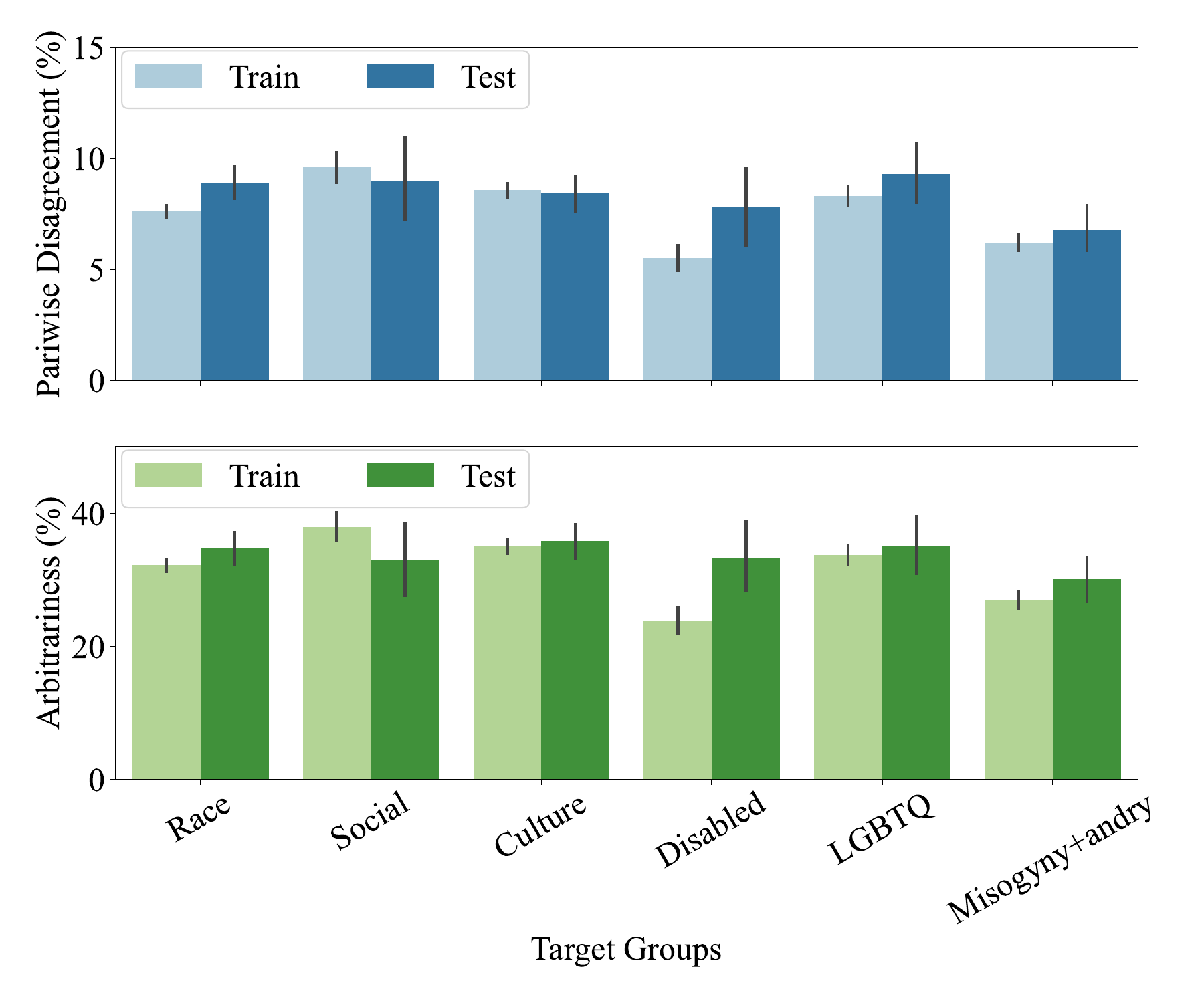}}
\subfloat[Jigsaw Fine-Tuned]{  \includegraphics[width=.4\linewidth]{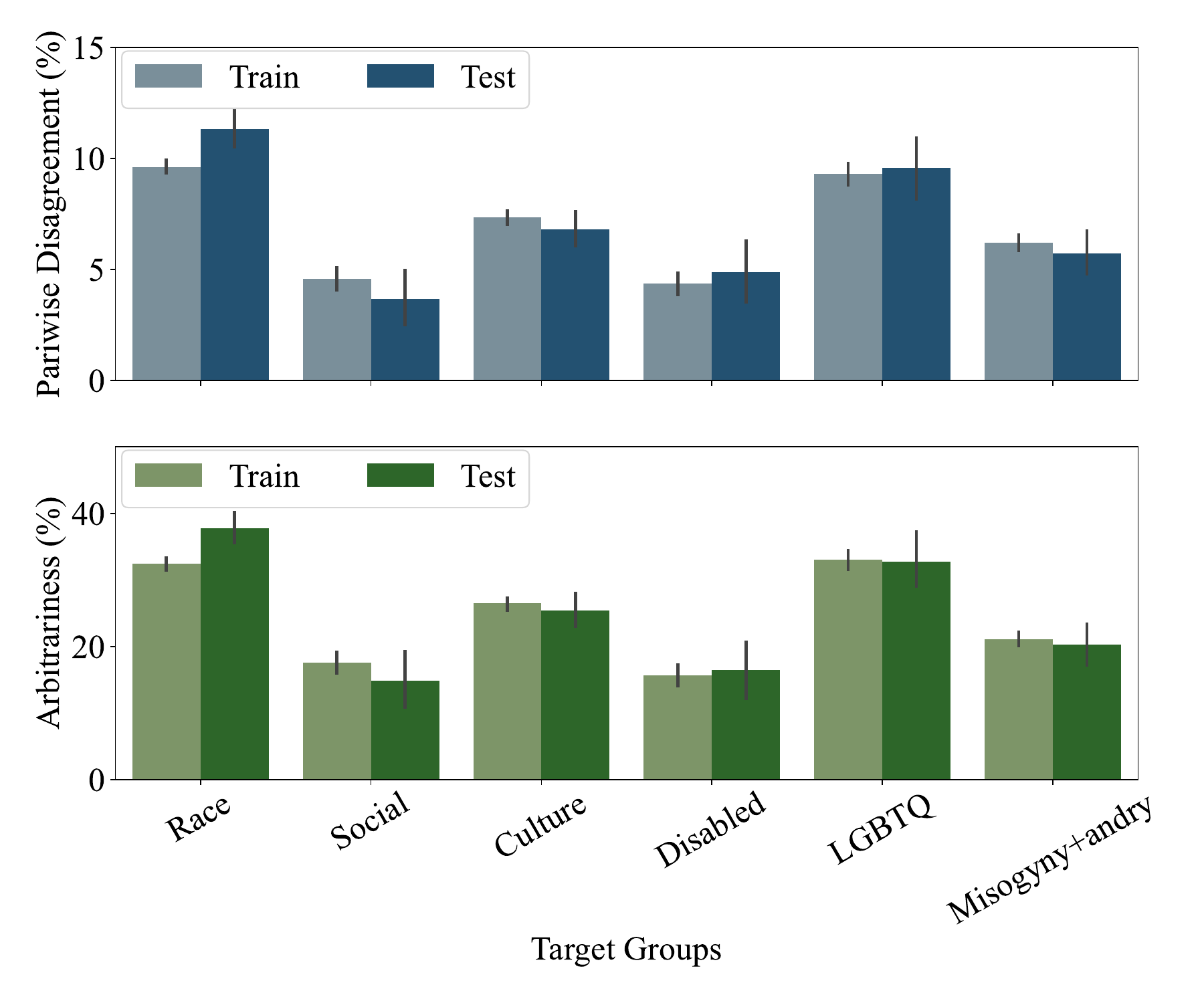}}
  \caption{Average pairwise disagreement and arbitrariness in different target groups for the fine-tuned Toxigen and Jigsaw models.
  The results show the pairwise disagreement in percentage (x-axis) for the union of four different datasets: DynaHate, SBF, Toxigen, and HateExplain.
  The results are shown for training and test partitions of each dataset.
  The confidence in the CP methods was chosen to be $50\%$ containing all fine-tuned models, leading to the selection of 38 out of 40 Roberta models in the Rashomon set fine-tuned in the Toxigen dataset and 17 out of 20 Jigsaw fine-tuned models.
  }
  \label{fig:appDemo50}
\end{figure}

\begin{figure}[!htbp]
  \centering
\subfloat[Toxigen Fine-Tuned]{
\includegraphics[width=.4\linewidth]{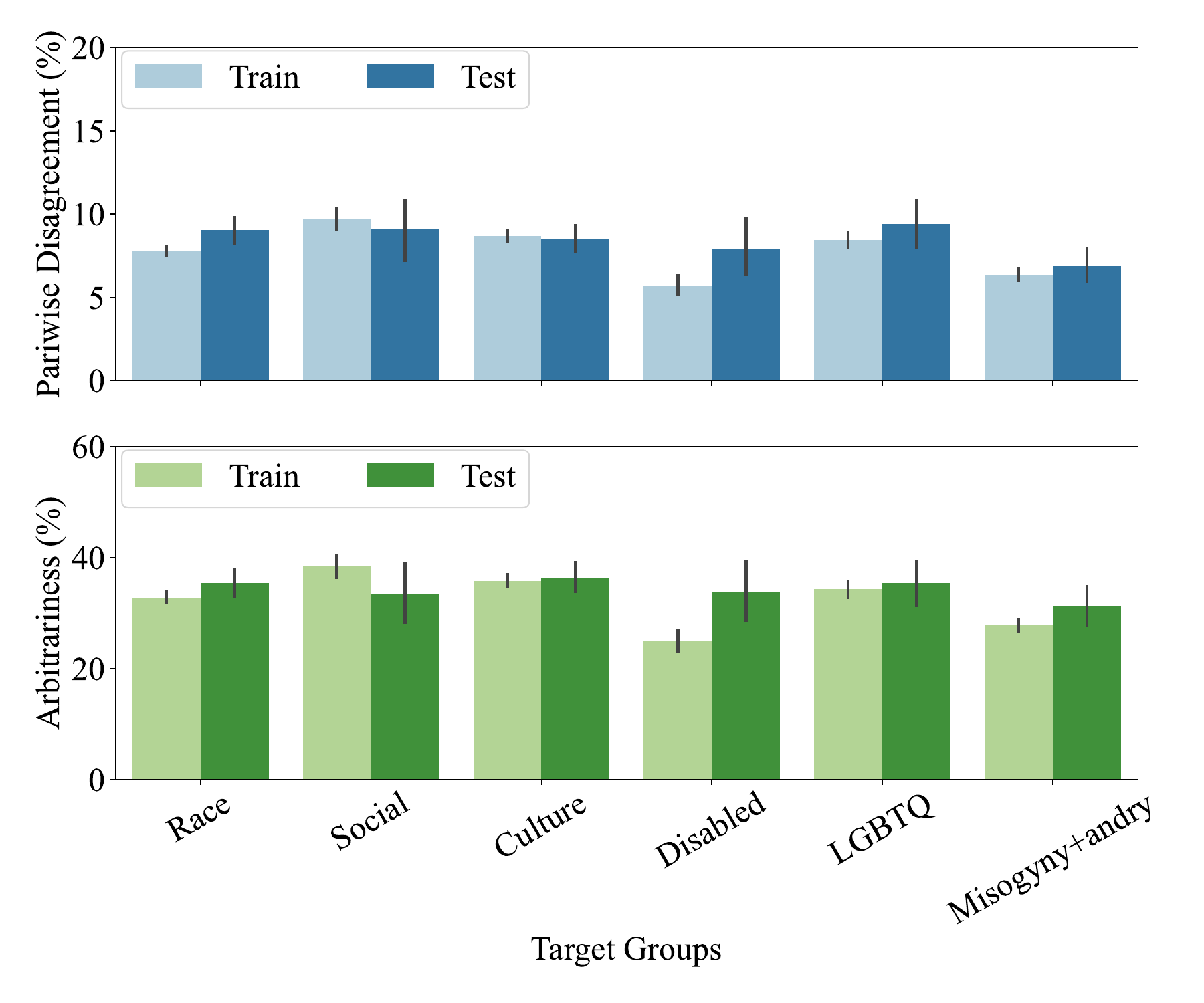}}
\subfloat[Jigsaw Fine-Tuned]{  \includegraphics[width=.4\linewidth]{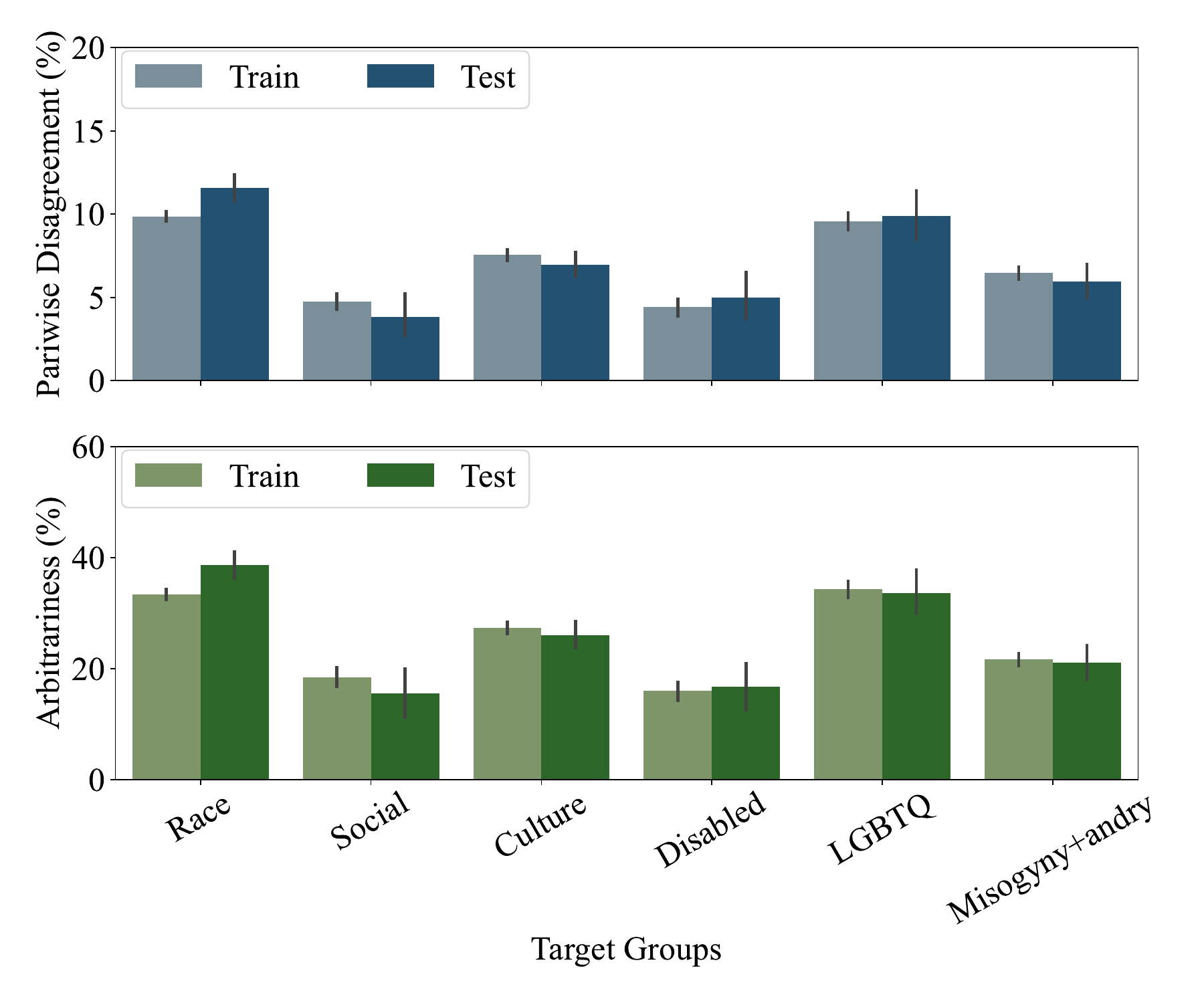}}
  \caption{Average pairwise disagreement and arbitrariness in different target groups for the fine-tuned Toxigen and Jigsaw models.
  The results show the pairwise disagreement in percentage (x-axis) for the union of four different datasets: DynaHate, SBF, Toxigen, and HateExplain.
  The results are shown for training and test partitions of each dataset.
  The confidence in the CP methods was chosen to be $1\%$ containing all fine-tuned models, leading to the selection of 40 out of 40 Roberta models in the Rashomon set fine-tuned in the Toxigen dataset and 20 out of 20 Jigsaw fine-tuned models.
  }
  \label{fig:appDemo1}
\end{figure}

Here, we also show how arbitrariness and pairwise disagreement vary across different targeted demographic groups.
Figures \ref{fig:appDemo50} and \ref{fig:appDemo1} indicate that even under higher confidence values, arbitrariness and disagreement are still non-uniformly distributed as showed in Figure \ref{fig:ArbiPerGroup}, leading to disparate algorithmic treatment.

\subsection{Human vs. Model arbitrariness}
We also display the arbitrariness and pairwise disagreement values across clear and unclear toxic content.
Recall that we consider \emph{clear} sentences the ones that all human annotators agreed upon its toxicity and \emph{unclear} when not all annotators classified the sentence toxicity equally. 

\begin{figure}[h]
  \centering
  \subfloat[Toxigen Fine-Tuned Models]{\includegraphics[width=.25\linewidth]{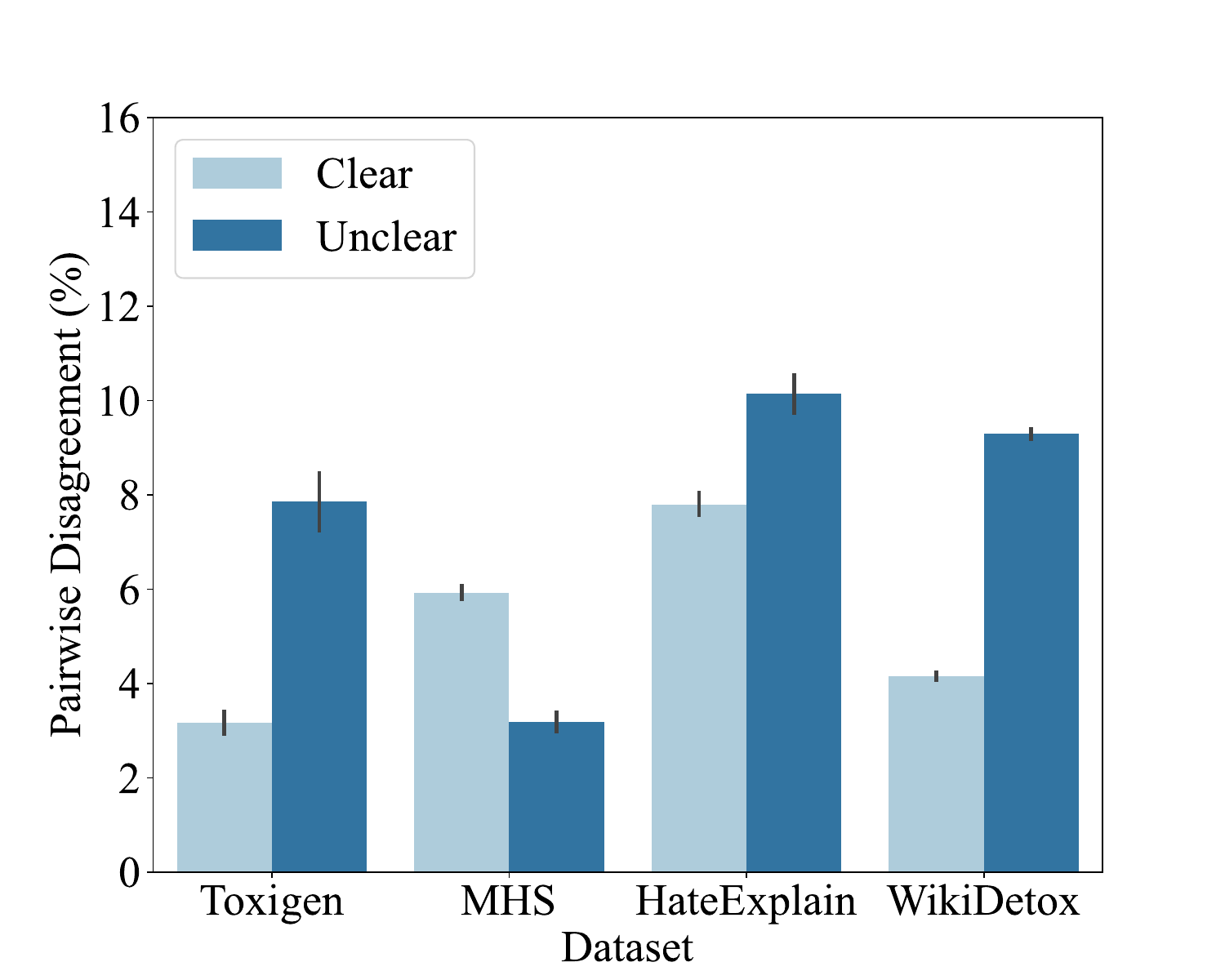}
  \includegraphics[width=.25\linewidth]{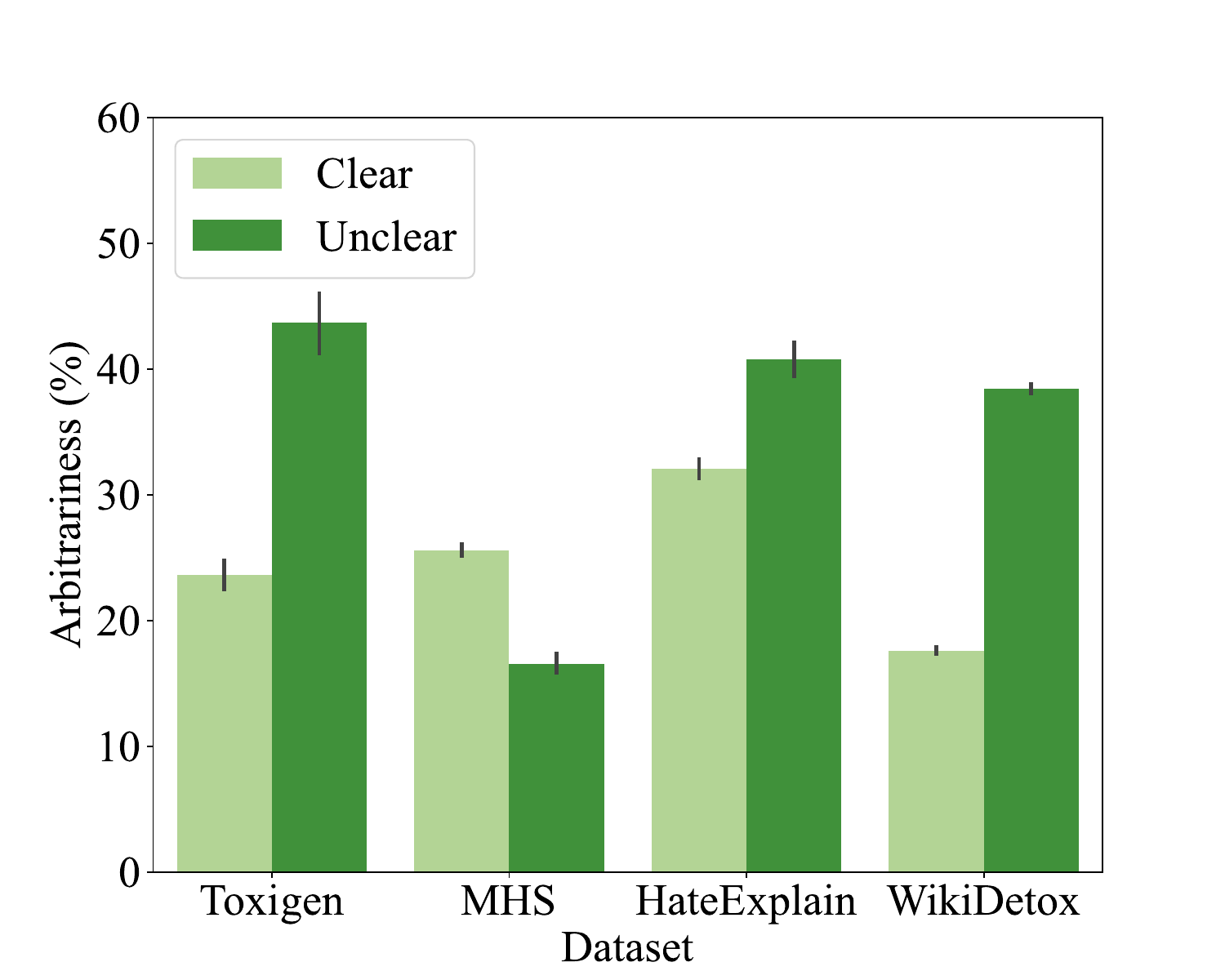}}
  \subfloat[Jigsaw Fine-Tuned Models]{\includegraphics[width=.25\linewidth]{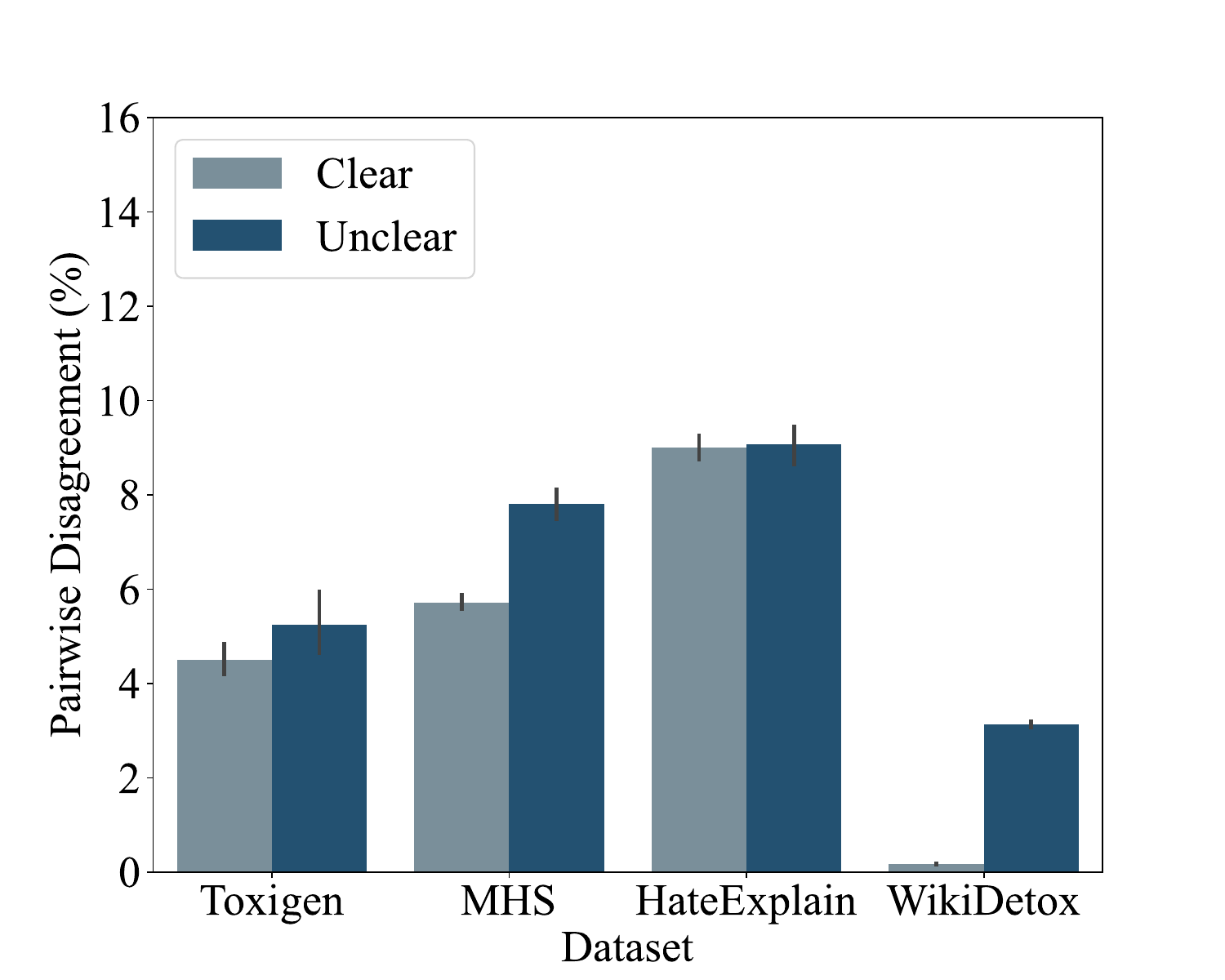}
  \includegraphics[width=.25\linewidth]{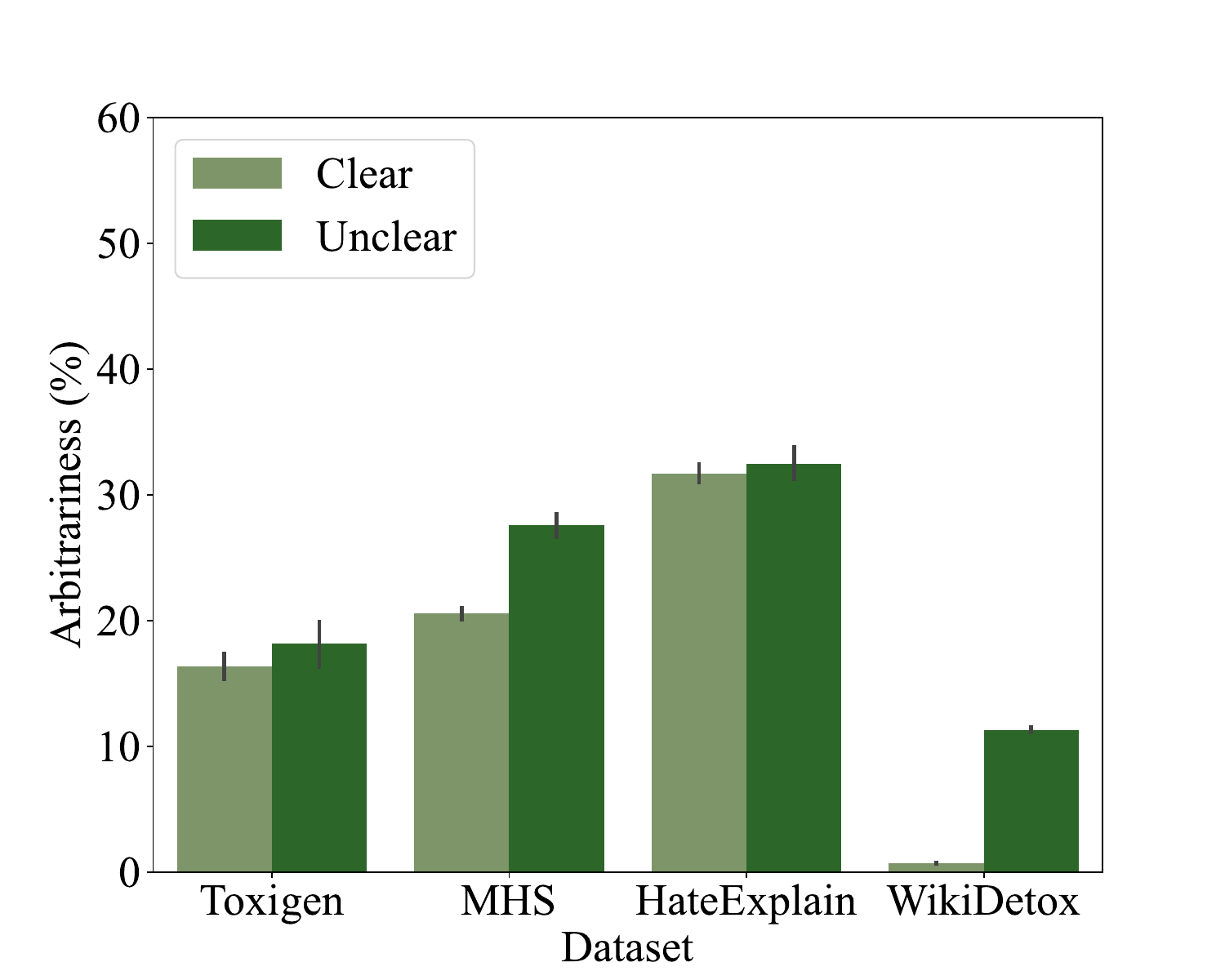}}
   \caption{Average pairwise disagreement and arbitrariness for \emph{Clear} and \emph{Unclear} sentences using the Toxigen fine-tuned and Jigsaw fine-tuned models. 
  The table shows the pairwise disagreement estimated values along with the $95\%$ confidence intervals using the standard error from the mean.
  We consider a sentence \emph{Unclear} when at least one annotator labeled the sentence differently than others and \emph{Clear} otherwise. 
  The confidence in the CP methods was chosen to be $1\%$, including all fine-tuned models in the above analysis.
  }
  \label{fig:appHumanMachine1}
\end{figure}

\begin{figure}[h]
  \centering
  \subfloat[Toxigen Fine-Tuned Models]{\includegraphics[width=.25\linewidth]{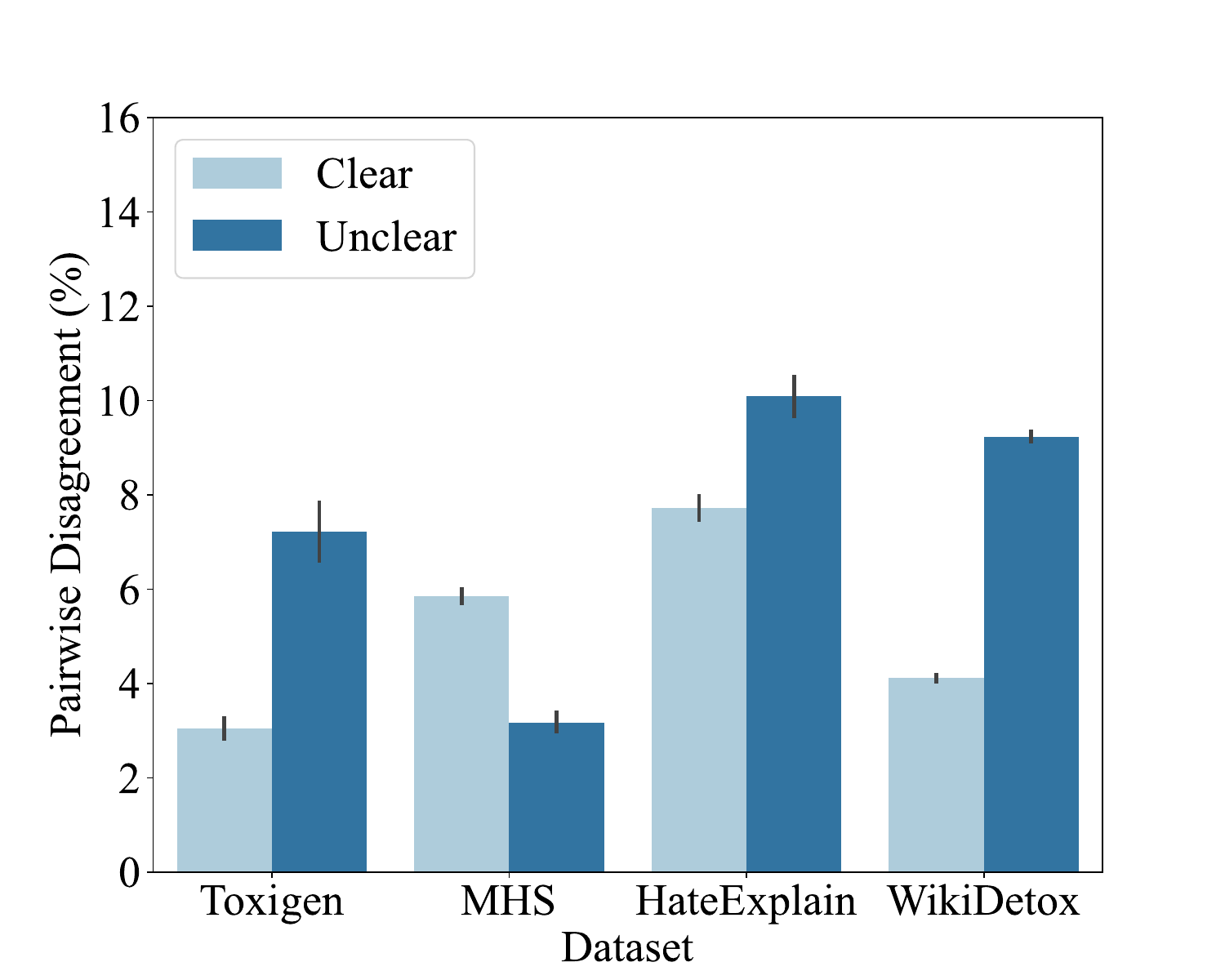}
  \includegraphics[width=.25\linewidth]{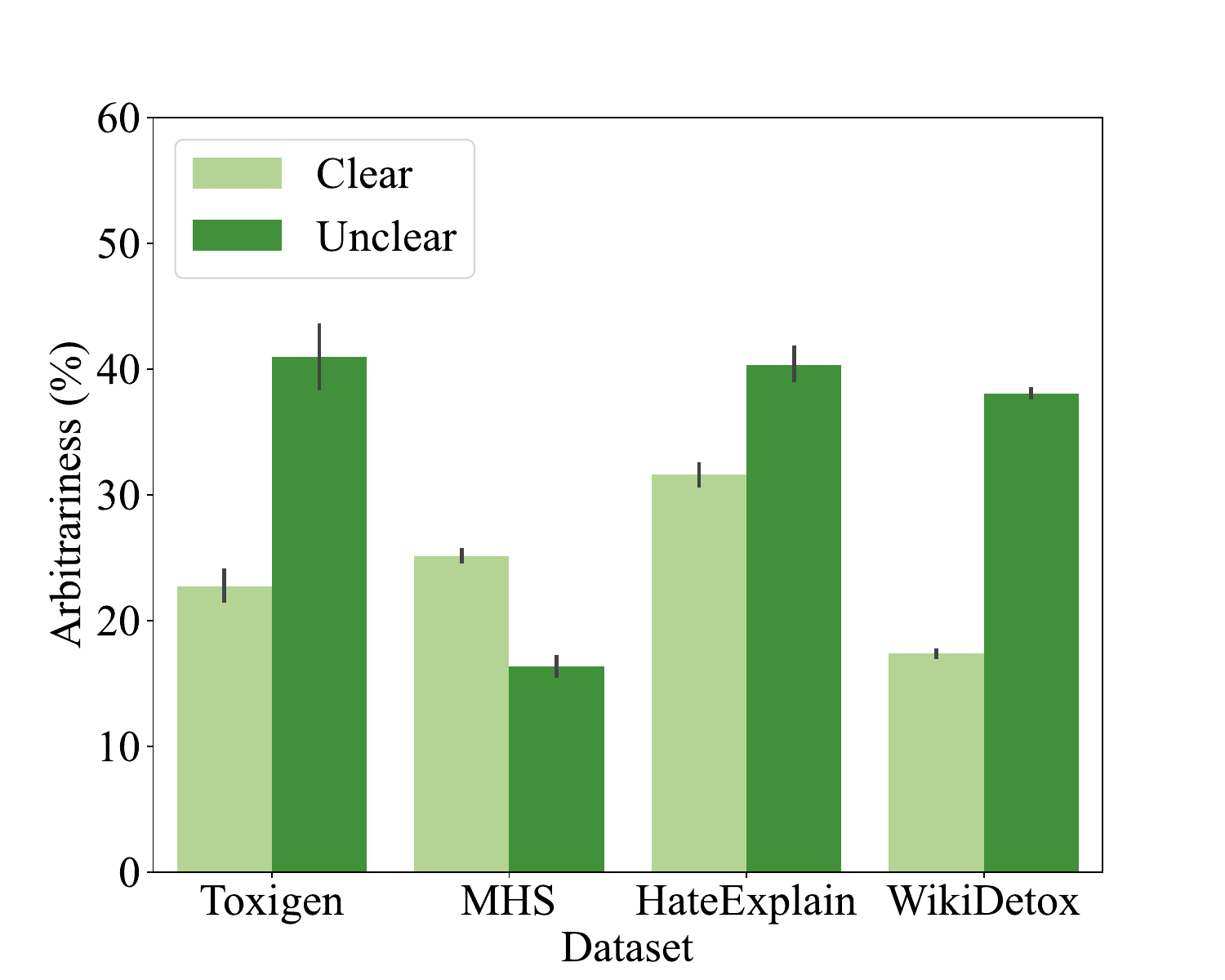}}
  \subfloat[Jigsaw Fine-Tuned Models]{\includegraphics[width=.25\linewidth]{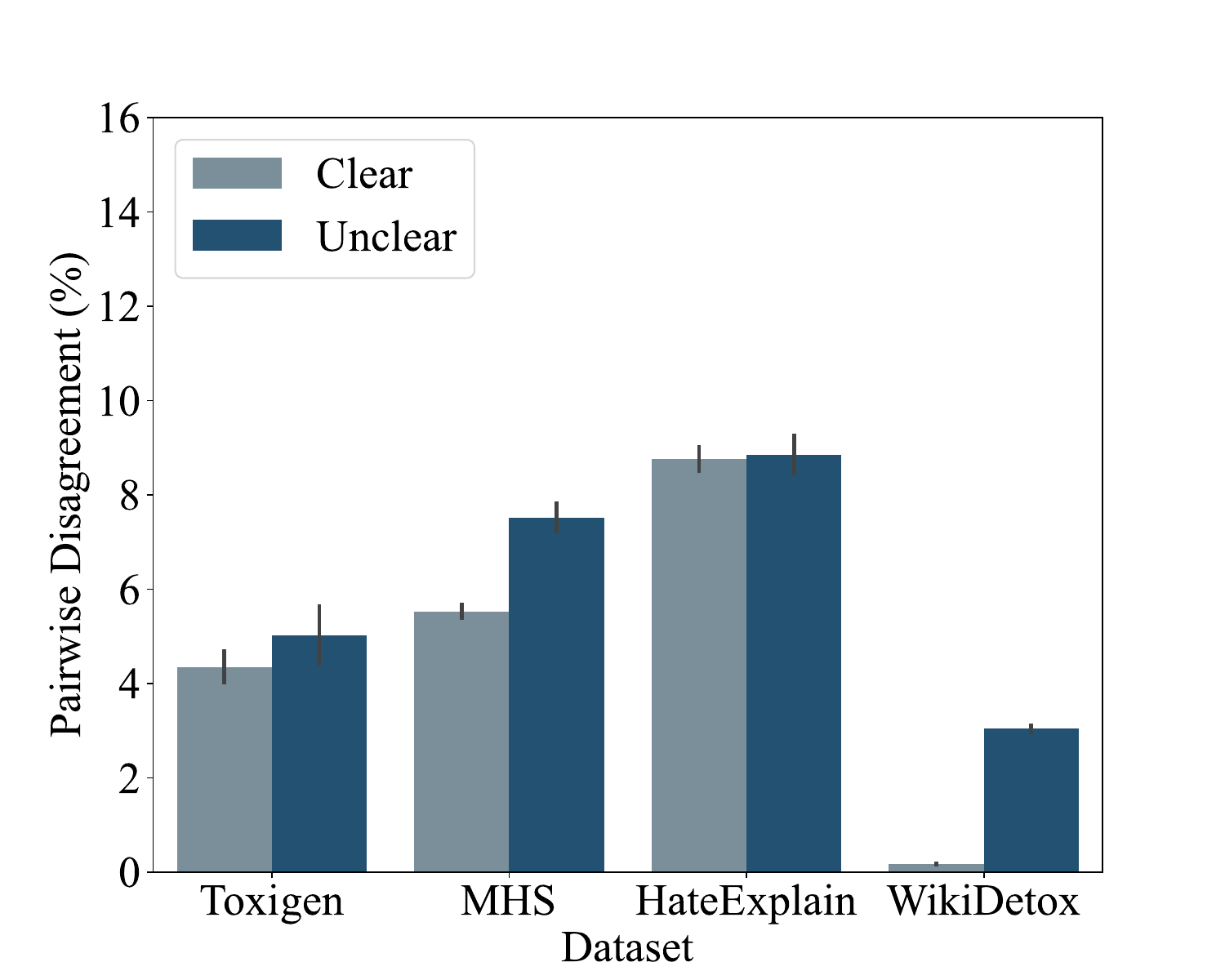}
  \includegraphics[width=.25\linewidth]{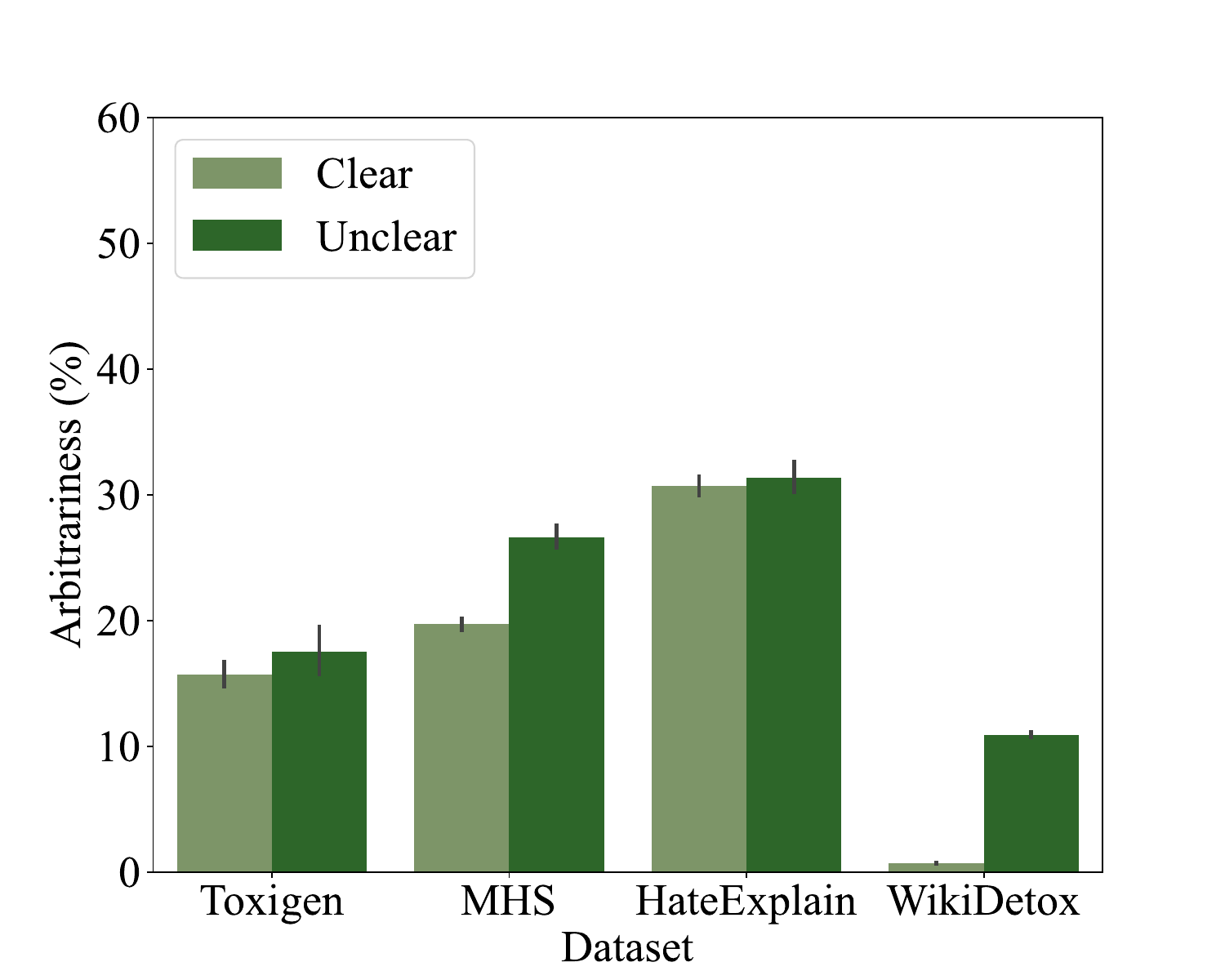}}
   \caption{Average pairwise disagreement and arbitrariness for \emph{Clear} and \emph{Unclear} sentences using the Toxigen fine-tuned and Jigsaw fine-tuned models. 
  The table shows the pairwise disagreement estimated values along with the $95\%$ confidence intervals using the standard error from the mean.
  We consider a sentence \emph{Unclear} when at least one annotator labeled the sentence differently than others and \emph{Clear} otherwise. 
  The confidence in the CP methods was chosen to be $50\%$, including all fine-tuned models in the above analysis.
  }
  \label{fig:appHumanMachine50}
\end{figure}

Figures \ref{fig:appHumanMachine1} and \ref{fig:appHumanMachine50} present the same pattern of higher arbitrariness and pairwise disagreement in unclear sentences while also having a high arbitrariness and pairwise disagreement in clear sentences --- and we discuss in Section \ref{sec:dataAnalysis}.

\end{document}